\documentclass[aps,pre,twocolumn,groupedaddress,floatfix,longbibliography,superscriptaddress]{revtex4-1}

\usepackage{amsmath}
\usepackage{amssymb}
\usepackage{amsfonts}
\usepackage{graphicx}
\usepackage{bbold}
\usepackage{bm}
\usepackage{bm}
\usepackage{cancel}
\usepackage{times,float}
\usepackage{graphicx}
\usepackage[usenames,dvipsnames,svgnames]{xcolor}
\usepackage{hyperref}
\hypersetup{colorlinks=true, linkcolor=Blue, citecolor=Blue,urlcolor=Blue}
\usepackage{multirow}
\usepackage{ulem}
\usepackage{color,epsfig}
\usepackage{bm}
\usepackage{slashed}
\usepackage{hyperref}
\usepackage{subfigure}
\usepackage{feynmf}
\usepackage{array}

\newcommand{\bea}{\begin{eqnarray}}
\newcommand{\eea}{\end{eqnarray}}
\newcommand{\be}{\begin{equation}}
\newcommand{\ee}{\end{equation}}

\newcommand{\nn}{\nonumber}
\newcommand{\ii}{\mathrm{i}}

\newcommand{\mbf}{\mathbf}
\newcommand{\mbb}{\mathbb}
\newcommand{\bs}{\boldsymbol}
\newcommand{\f}{\phi}
\newcommand{\p}{\pi}

\newcommand{\lb}{\lambda}
\newcolumntype{P}[1]{>{\centering\arraybackslash}p{#1}}
\newcommand{\Tr}{\text{Tr}}
\newcommand{\laplace}[1]{\mathcal{L}\left\{#1\right\}}

\begin{document}

\title{Electromagnetic coupling and transport in a topological insulator-graphene hetero-structure}
\author{Daniel A. Bonilla}
\email{dabonilla@uc.cl}
\affiliation{Facultad de F\'isica, Pontificia Universidad Cat\'olica de Chile, Vicu\~{n}a Mackenna 4860, Santiago, Chile}
\author{Jorge David Casta\~no-Yepes}
\email{jcastano@uc.cl}
\affiliation{Facultad de F\'isica, Pontificia Universidad Cat\'olica de Chile, Vicu\~{n}a Mackenna 4860, Santiago, Chile}
\author{A. Mart\'in-Ruiz}
\email{alberto.martin@nucleares.unam.mx}
\address{Instituto de Ciencias Nucleares, Universidad Nacional Aut\'{o}noma de M\'{e}xico, 04510 Ciudad de M\'{e}xico, M\'{e}xico}
\author{Enrique Mu\~noz}
\email{munozt@fis.puc.cl}
\affiliation{Facultad de F\'isica, Pontificia Universidad Cat\'olica de Chile, Vicu\~{n}a Mackenna 4860, Santiago, Chile}
\affiliation{Center for Nanotechnology and Advanced Materials CIEN-UC, Avenida Vicuña Mackenna 4860, Santiago, Chile}

\begin{abstract}
The electromagnetic coupling between hetero-structures made of different materials is of great interest, both from the perspective of discovering new phenomena, as well as for its potential applications in novel devices. In this work, we study the electromagnetic coupling of a hetero-structure made of a topological insulator (TI) slab and a single graphene layer, where the later presents a diluted concentration of ionized impurities. We explore the topological effects of the magneto-electric polarizability (MEP) of the TI, as well as its relative dielectric permittivity on the electrical conductivity in graphene at low but finite temperatures. 
\end{abstract}

\maketitle

\section{Introduction}

 As stated by the great Aristotle more than 2000 years ago in his work Metaphysics\cite{Aristotle} “{\it{To return to the difficulty which has been stated with respect both to definitions and to numbers, what is the cause of their unity? In the case of all things which have several parts and in which the totality is not, as it were, a mere heap, but the whole is something beside the parts, there is a cause; for even in bodies contact is the cause of unity in some cases, and in others viscosity or some other such quality.}}" The combination of different materials in the form of hetero-structures\cite{Geim2013} is our modern quest to search for novel properties that emerge beyond the trivial superposition of those of their individual parts. This search is of great interest not only from a fundamental perspective, since exciting new phenomena may be observed, but also to engineer materials for applications in novel devices. Among the different emerging phenomena in hetero-structures, electromagnetic effects are highly relevant for the transmission and storage of energy and information. In this context, the control of electronic transport properties is of fundamental importance.

 The discovery of novel materials with non-trivial topological properties~\cite{Chiu_2016}, such as topological insulators (TIs)~\cite{Hasan_2010}, Dirac and Weyl semimetals~\cite{Hasan_2017}, has introduced a plethora of new phenomenology. In particular, the existence of gapless edge (in 2D TIs) or surface (in 3D TIs) pseudo-relativistic chiral states~\cite{Roushan_2009} makes them excellent potential candidates for applications in quantum information technologies and thermoelectrics~\cite{Xu_2017}. In addition, the so-called magneto-electric polarizability (MEP)~\cite{PhysRevB.78.195424,Fiebig_2005} that locally modifies the constitutive relations between the electromagnetic fields in TIs, provides new opportunities to control the magnetoelectric response in such systems. Even though these effects have been extensively studied in individual topological materials, their electromagnetic coupling when integrated into hetero-structures remains a vast territory for further exploration~\cite{CASTANOYEPES2020114202,castano2022ground}. 
 
\begin{figure}[h!]
    \centering
    \includegraphics[scale=0.55]{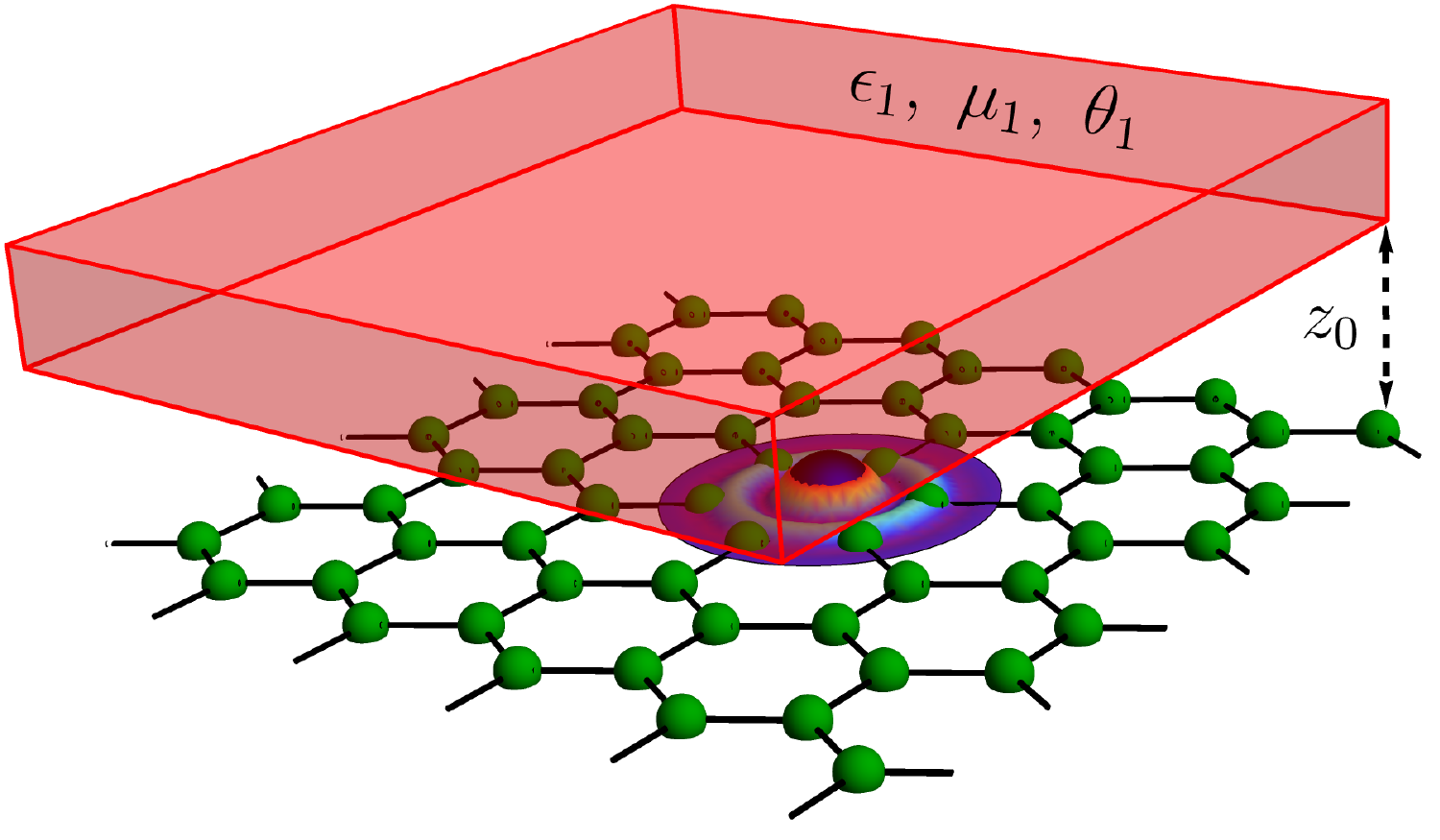}
    \caption{Pictorial representation of the system. A hetero-structure composed by a TI slab (with material properties $\epsilon_1,\mu_1$, and $\theta_1$) and a graphene monolayer. The two materials are separated by a distance $z_0$. A diluted concentration of ionized impurities is present in the graphene monolayer. }
    \label{fig:scheme_ion}
\end{figure}

 In this work, we consider a hetero-structure composed of a TI slab and a single graphene layer, as depicted in Fig.~\ref{fig:scheme_ion}. We further assume that a diluted concentration of ionized impurities is present in the graphene monolayer. The presence of such charged impurities will induce a local distortion of the charge density of the 2D electron gas, leading to a non-trivial electromagnetic coupling between the TI and the graphene monolayer in the heterostructure. As a probe of this coupling, we further studied the electrical conductivity as a function of temperature, by including the scattering effects with the local electromagnetic field configuration via the Kubo linear response formalism~\cite{nano12203711,Falomir_2021,Juricic_Munoz_Soto_2022}. We applied our theoretical results to model the electromagnetic coupling in hetero-structures made of different TIs (PbTe, Bi$_2$Te$_3$, PbSe, PbS, Bi$_2$Se$_3$, TlBiSe$_2$, TbPO$_4$).
 Our analytical and numerical results suggest that, among the properties of the TIs, the dielectric permittivity $\epsilon_1$ is the most relevant at tuning the electronic transport in the coupled graphene monolayer. On the other hand, we also observed that the topological effects arising from the presence of the MEP coefficient $\theta$ are comparatively very small even at zero temperature.

\section{Electromagnetic response of the TI} \label{EM_Resp_TI}

The effective field theory governing the electromagnetic response of topological insulators, independently of the microscopic details, is defined by the action (in SI units) \cite{PhysRevB.78.195424}:
\begin{align}
    \mathcal{S} = \int d^{4}x \, \Bigg\{\ \!\!  \frac{1}{2} \left[  \epsilon \, {\bf{E}} ^{2} - ( 1 / \mu ) \, {\bf{B}} ^{2} \right] + \frac{\alpha}{\pi} \sqrt{\frac{\epsilon _{0}}{\mu _{0}}} \theta \, {\bf{E}} \cdot {\bf{B}} \Bigg\}\ , \label{Action}
\end{align}
where $\alpha = e^{2}/ ( 4 \pi \epsilon _{0} \hbar c ) \simeq 1/137$ is the fine structure constant, $\epsilon$ and $\mu$ are the permittivity and permeability of the material, respectively, and $\theta$ is the topological magnetoelectric polarizability (MEP) or axion field. The coupling between the gauge field and the free sources is introduced as usual. TR symmetry indicates that $\theta = 0, \pi $ (mod $2 \pi$), and hence the $\theta$-term in the action of Eq.~(\ref{Action}) has no effect on Maxwell equations in the bulk. The nontrivial topological property, a surface half-integer quantum Hall effect, manifests only when a TR-breaking perturbation is induced on the surface to gap the surface states, thereby converting the material into a full insulator. This can be achieved by introducing magnetic dopants to the surface \cite{doi:10.1126/science.1230905} or by the application of an external static magnetic field \cite{PhysRevLett.105.166803}. In this situation, $\theta$ is quantized in odd integer values of $\pi$, where the magnitude and sign of the multiple is controlled by  the strength and direction of the TR-breaking perturbation.

The field equations arising from the theory of Eq.~(\ref{Action}) are those of Maxwell electrodynamics in a medium with the modified constitutive relations \cite{PhysRevB.78.195424}
\begin{align}
{\bf{D}} = \epsilon {\bf{E}} + \alpha \frac{\theta}{\pi} \,  \sqrt{\frac{\epsilon _{0}}{\mu _{0}}}  \,  {\bf{B}} ,  \qquad {\bf{H}} = \frac{1}{\mu} {\bf{B}} - \alpha \frac{\theta}{\pi} \,  \sqrt{\frac{\epsilon _{0}}{\mu _{0}}}  \,  {\bf{E}}  . \label{ConstEquations}
\end{align}

The $\theta$-dependent term in each constitutive equation encodes the most salient feature of this theory: the topological magneto-electric effect, where an electric field can induce a magnetic polarization and a magnetic field can induce an electric polarization \cite{Fiebig_2005}.


The general solution to the field equations can be expressed in terms of an indexed Green's function $G_{\mu \nu}$ which satisfies the field equations for a point-like source and appropriate boundary conditions, namely,
\begin{align}
    A^\mu(\mathbf{r})=\int _{V}  G^\mu_{~\nu}(\mathbf{r},\mathbf{r}') \, J ^\nu(\mathbf{r}') \, d^{3} \mathbf{r}' ,
\end{align}
where $\mathbf{r}$ and $\mathbf{r}'$ are the coordinates of the field-observation and the source, respectively. Here, $A^{\mu} = (\Phi / c , {\bf{A}})$ is the four-potential and $J ^{\mu} = (\rho c , {\bf{J}})$ is the four-current density. The exact form of the indexed Green function depends on the geometry configuration of the problem. For example, explicit expressions for the problem of two topologically insulating media separated by a planar interface can be found in Refs.~\cite{PhysRevD.92.125015, PhysRevD.93.045022}. The corresponding expressions for a spherical interface is reported in Refs.~\cite{PhysRevD.94.085019, CASTANOYEPES2020114202}, and the results for a cylindrial interface have been reported in Ref.~\cite{PhysRevD.98.056012}. The full expressions are not illuminating at all, so here we just concentrate in the problem at hand, where the source is purely electric, i.e. we take $J^{i}=0$.

\begin{figure}
    \centering
    \includegraphics[scale=0.9]{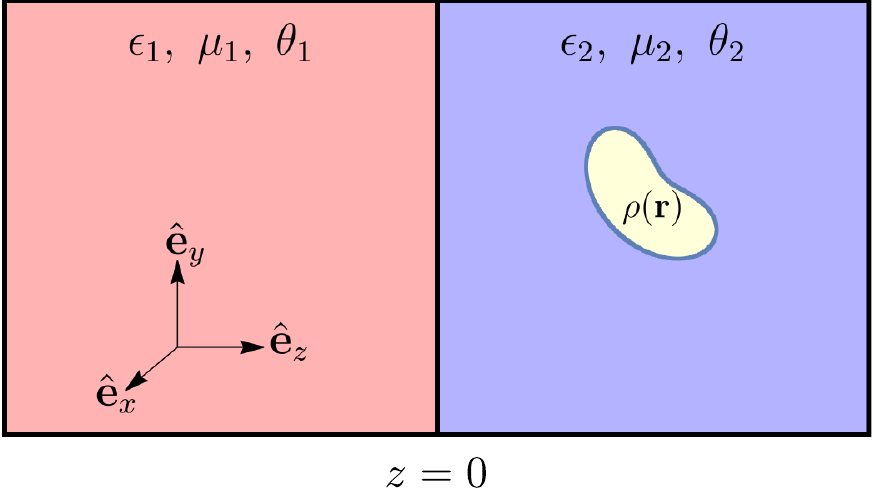}
    \caption{Charge density $\rho(\mathbf{r})$ near to a planar surface}
    \label{Fig2}
\end{figure}

Let us consider the particular configuration depicted in Fig. \ref{Fig2}. The half-space $z<0$ is occupied by a topological insulator with a dielectric constant $\epsilon _{1}$, a magnetic permeability $\mu _{1}$,  and MEP $\theta _{1}$,  while the half-space $z>0$ is occupied by a material (topologically trivial or not) with a dielectric constant $\epsilon _{2}$,  a magnetic permeability $\mu _{2}$, and MEP $\theta _{2}$. A charge distribution $\rho ({\bf{r}})$ is placed in the region $z>0$, as shown in Fig. \ref{Fig2}. The electric potential becomes
\begin{align}
    \Phi ({\bf{r}}) = \int \mathcal{G} ({\bf{r}} , {\bf{r}}^{\prime}) \, \rho ({\bf{r}}^{\prime}) \, d^{3} {\bf{r}}^{\prime} ,  \label{ElectricPot}
\end{align}
where the corresponding Green function (i.e. the $00$-component of the indexed Green's function), for $z>0$, is
\begin{align}
    \mathcal{G} ({\bf{r}},{\bf{r}}^{\prime}) \equiv G^{0}_{~ 0}(\mathbf{r},\mathbf{r}^{\prime}) = \frac{1}{4 \pi \epsilon _{2}} \left( \frac{1}{r _{+}} + \frac{\kappa}{r _{-}}\right) , \label{G00-GF}
\end{align}
with $r_{\pm} = \sqrt{(x-x^{\prime})^{2} + (y-y^{\prime})^{2} + (z\mp z^{\prime})^{2} }$ and
\begin{align}
    \kappa \equiv \frac{ \epsilon _{2} - \epsilon _{1} - \Delta }{ \epsilon _{2} + \epsilon _{1} + \Delta } , \qquad \Delta \equiv \frac{\mu _{1} \mu _{2}}{\mu _{1} + \mu _{2}} \left(  \alpha \frac{\theta _{1} - \theta _{2}}{\pi} \sqrt{\frac{\epsilon _{0}}{\mu _{0}}} \right) ^{2} . 
    \label{kappa_and_g}
\end{align}

The magnetic response, which is purely topological since the source is electric, is determined by the $0i$-components of the indexed Green's function. One obtains
\begin{align}
    {\bf{A}} ({\bf{r}}) = \int {\bf{G}} ({\bf{r}} , {\bf{r}}^{\prime}) \, \rho ({\bf{r}}^{\prime}) \, d^{3} {\bf{r}}^{\prime} ,  \label{VectPotential}
\end{align}
where the corresponding Green's vector, for $z>0$, is
\begin{align}
    {\bf{G}} ({\bf{r}} , {\bf{r}}^{\prime}) \equiv G^{i}_{~ 0} (\mathbf{r},\mathbf{r}^{\prime}) \, \hat{{\bf{e}}}_{i} = \frac{\mu _{1} g}{4 \pi} \, \frac{\hat{{\bf{e}}}_{z} \times {\boldsymbol{\rho}} }{\rho ^{2}} \left( 1 - \frac{z+z^{\prime}}{r_{-}} \right) , \label{Green_Vector}
\end{align}
with ${\boldsymbol{\rho}} = (x-x^{\prime}) \hat{{\bf{e}}}_{x} + (y-y^{\prime}) \hat{{\bf{e}}}_{y} $ and
\begin{align}
    g = \alpha \frac{\theta _{1} - \theta _{2}}{\pi} \sqrt{\frac{\epsilon _{0}}{\mu _{0}}} \,  \frac{\mu _{2}}{\mu _{1} + \mu _{2}} \frac{2}{ \epsilon _{1} + \epsilon _{2} + \Delta }  .
\end{align}

As expected, in the non-topological limit $\theta _{1} \to \theta _{2}$ we get $\Delta =0$ and $g=0$, since the topological magnetoelectricity disappears. Besides, $\kappa \to (\epsilon _{2} - \epsilon _{1}) / (\epsilon _{2} + \epsilon _{1}) $, which is the usual electrostatic result. As a consistency check one can further verify the image magnetic monopole effect of topological insulators \cite{doi:10.1126/science.1167747}: for a poinlike charge of strength $q$ at $z_{0}$, the electric field can be interpreted in terms of the original charge plus an image charge of strength $\kappa q$ at $- z_{0}$, and the magnetic field can be interpreted as that of a magnetic monopole of strength $q g$ at $- z_{0}$.

In the next section we apply the above results to the electronic density distribution that represents the physical configuration depicted in Fig.~\ref{fig:scheme_ion}.

\section{Electronic response due to the ionized impurity in the Graphene monolayer}
{In this section, we shall present the effective continuum model, in coordinate space, to account for the electromagnetic effects of a single charged impurity in the monolayer graphene located at a distance $z_{0}$ from the surface of a planar topological insulator, as shown in Fig.~\ref{fig:scheme_ion}. Let us assume that the unperturbed uniform electron gas density in the graphene monolayer is $\rho_0 = (-e)n_c$, with $n_c = N/A$ the free carrier density. If the ionized impurity has a charge $Q$, and is localized at the point $\mathbf{r}_0$, it will contribute to the total charge density with a delta distribution $Q\delta(\mathbf{r} - \mathbf{r}_0)$. In response to this charge, the 2D electron gas in the graphene monolayer will redistribute itself in order to screen it, leading to a small local deviation from the uniform charge density $\rho(\mathbf{r}) - \rho_0 \equiv \rho_{Y}(\mathbf{r})$. After a standard many-body treatment of the 2D electron gas via the random phase approximation (RPA), such screening can be well captured in the static regime via the Thomas-Fermi model~\cite{DasSarma_011}, that leads to a Yukawa density distribution in coordinate space,
\bea
\rho _\text{Y} (\mathbf{r})=Q\delta(\mathbf{r}-\mathbf{r}_0)-\frac{Q}{2 \pi l_0}\frac{e^{-|\mathbf{r}-\mathbf{r}_0|/l_0}}{|\mathbf{r}-\mathbf{r}_0|}\delta(z-z_0). \label{SourceYukawa}
\eea

Here, the inverse screening length $l_0^{-1}$ is given by the Thomas-Fermi wave-vector $q_\text{TF}$, defined by~\cite{DasSarma_011}
\bea
 q_\text{TF} = \frac{2\pi e^2}{\epsilon_2}D(\epsilon_F),
\label{eq:qTF}
\eea
where we have defined the density of states at the Fermi level $\epsilon_\text{F} = \hbar v_\text{F} k_\text{F}$,
\bea
D(\epsilon_\text{F}) = 4 \int \frac{d^2 k}{(2\pi)^2}\delta\left(\hbar v_\text{F} k - \epsilon_\text{F} \right) = \frac{2}{\pi}\frac{\epsilon_F}{\left( \hbar v_\text{F} k_\text{F}\right)^2}.
\label{eq:DE}
\eea

Finally, by using the definition of the Fermi wave-vector in mono-layer graphene as a function of the free carrier density $n_c$
\bea
k_\text{F} = \sqrt{\pi n_c},
\label{eq:kF}
\eea
we obtain from Eq. \eqref{eq:qTF} and Eq. \eqref{eq:DE} that the Yukawa-screening length is given by
\bea
l_0^{-1}\equiv q_\text{TF} = 4\frac{e^2}{\epsilon_2}\frac{\sqrt{\pi\,n_c}}{\hbar v_\text{F}}.
\label{eq:qTF2}
\eea
}

As discussed in the previous section, the presence of this local deviation $\rho_{Y}(\mathbf{r})$ in the charge density will act as a source to generate an electromagnetic response at the TI, in the form of scalar $\Phi(\mathbf{r})$ and vector potentials $\mathbf{A}(\mathbf{r})$, respectively. These electromagnetic potentials will exist not only at the TI itself, but also at the graphene monolayer, thus generating an electromagnetic coupling between the two materials that constitute the hetero-structure.

\begin{figure*}
    \centering
    \includegraphics[scale=0.5]{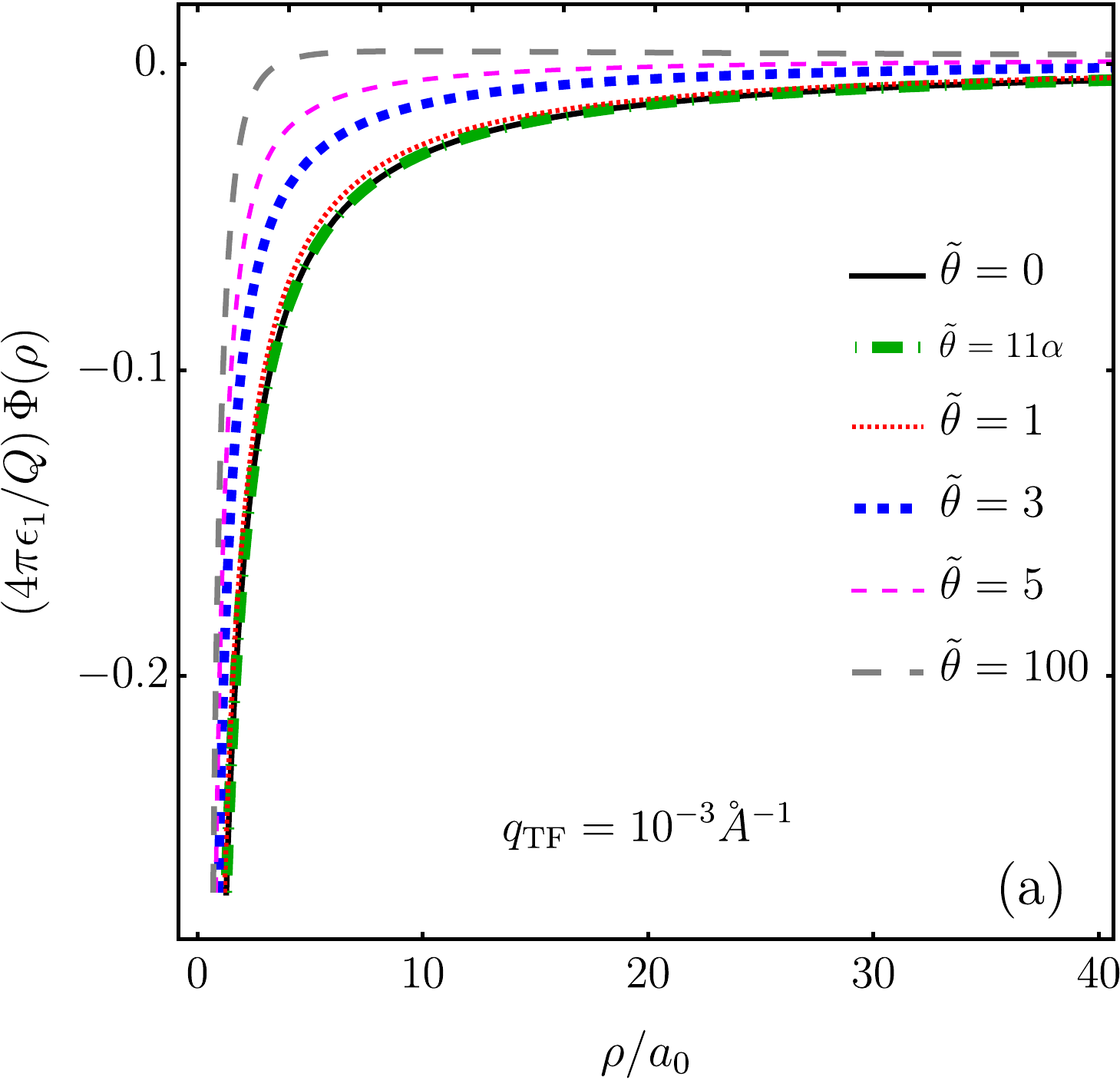}\hspace{1cm}\includegraphics[scale=0.5]{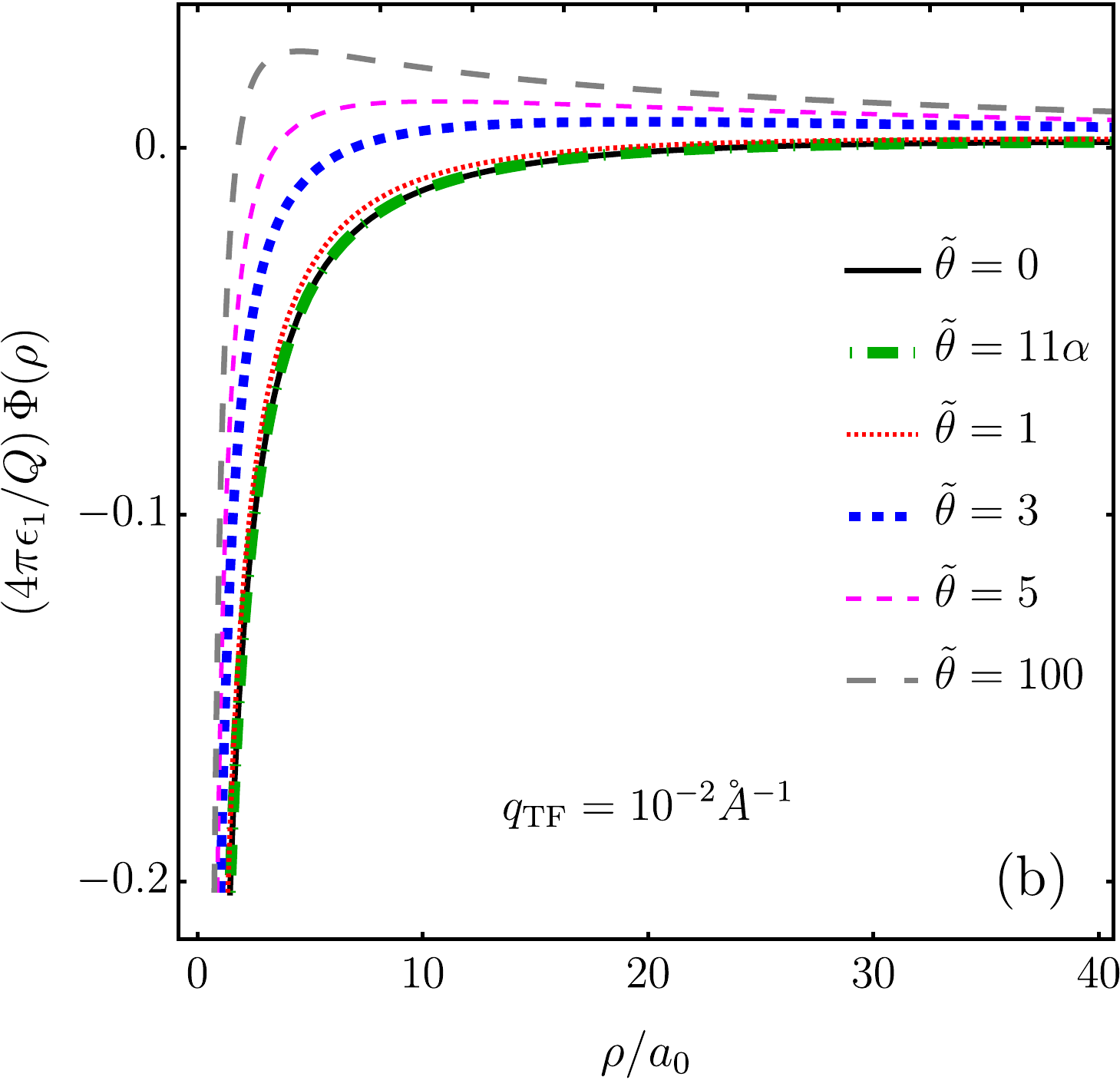}\\
    \vspace{0.5cm}
    \hspace{0.3cm}\includegraphics[scale=0.5]{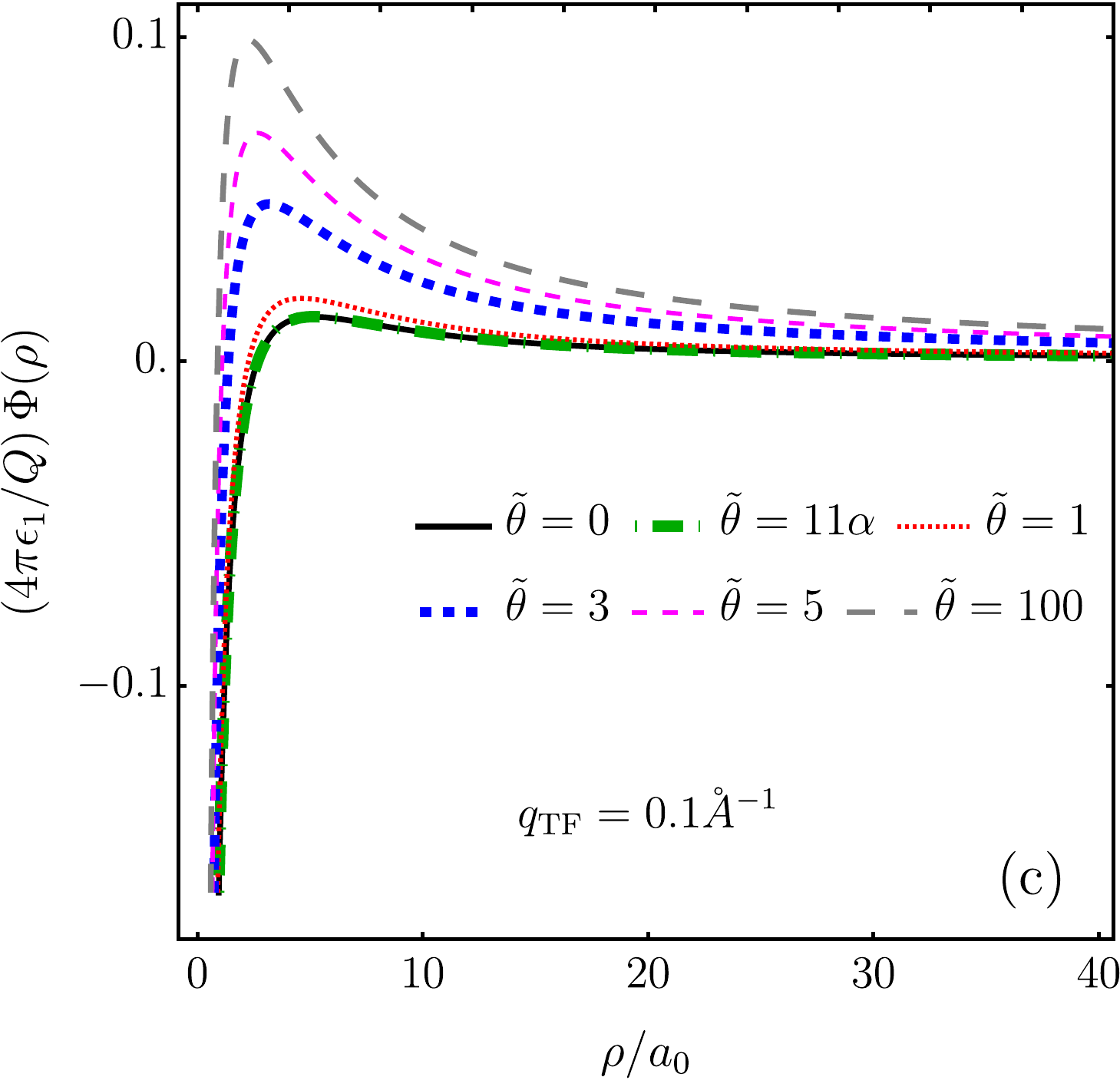}\hspace{1cm}\includegraphics[scale=0.5]{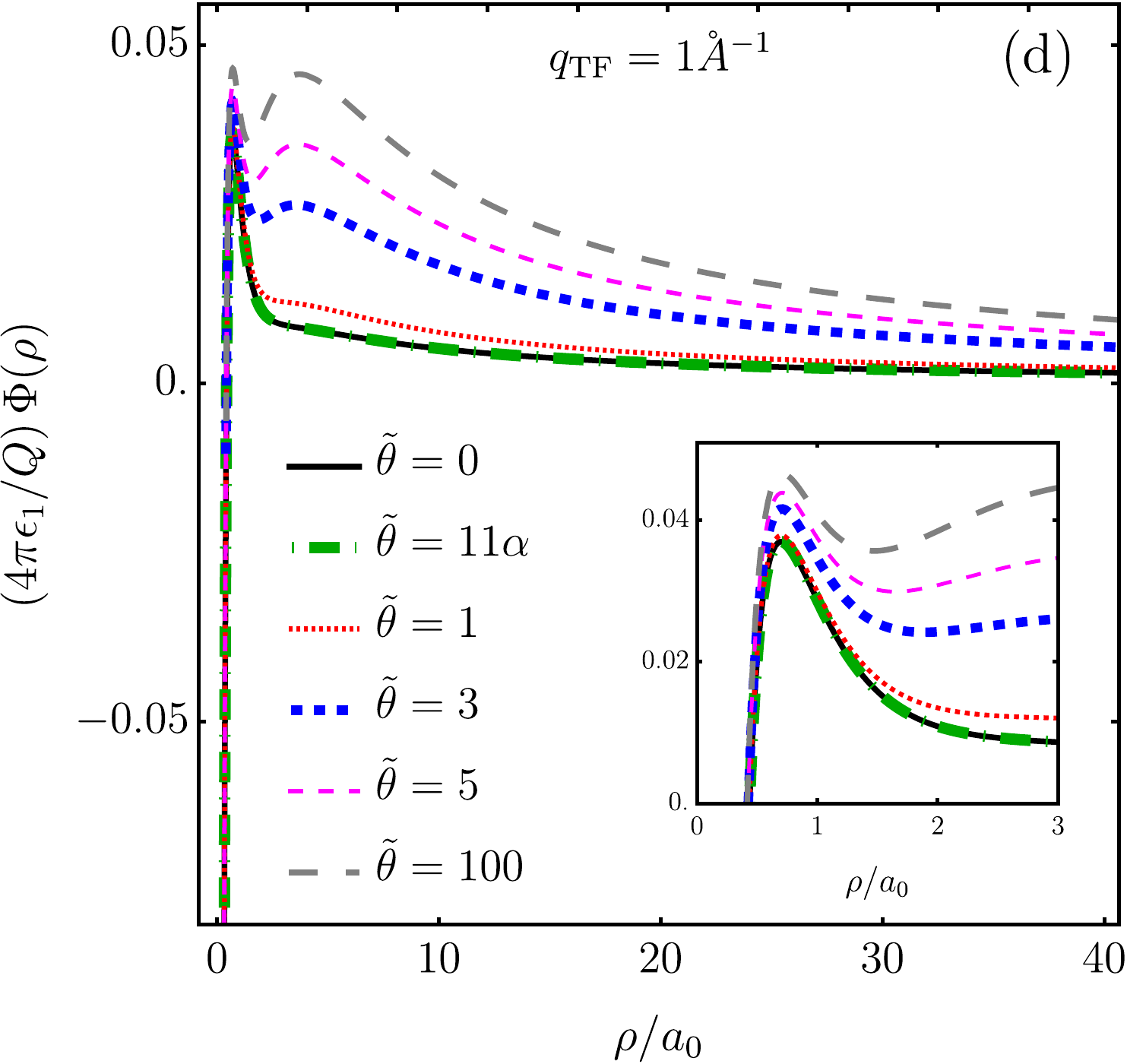}
    \caption{Scalar potential of Eq.~\eqref{Eq:Scalarpotentialfinal} with $z_0=1.42\AA$, as a function of $\rho/a_0$, where $a_0=2.46\AA$ is the lattice constant of the graphene. The panels are constructed for several values of $q_\text{TF}$ and $\widetilde{\theta}$, and we assumed $z_0=1.42\AA$.}    \label{fig:potential_theta_tildes}
\end{figure*}

To compute the electromagnetic potentials we use the theory discussed in the previous Section. We leave the details of the technical calculations to the Appendix~\ref{ap:scalarpotential} and we present here only the final results. For definiteness we evaluate the potentials at $z=z_{0}$, which is the position of the graphene monolayer as measured from the TI surface. So, we take ${\bf{r}} = \boldsymbol{\rho} + z _{0} \hat{{\bf{e}}}_{z}$. We first compute the scalar potential $\Phi(\mathbf{r})$, which is given by Eq. (\ref{ElectricPot}) with the Green's function given by Eq. (\ref{G00-GF}), with the charge density $\rho _\text{Y} (\mathbf{r})$ of Eq. (\ref{SourceYukawa}). The final result is
\begin{align}
    \Phi (\boldsymbol{\rho}) = \frac{Q}{4 \pi \epsilon _{1}} \left[ \frac{1}{\rho} + \frac{\kappa}{\sqrt{\rho ^{2} + (2 z_{0})^{2} }} - \Lambda ^{(0 )} _{0} (\rho) - \kappa  \Lambda ^{(0 )} _{1} (\rho) \right] , \label{Eq:Scalarpotentialfinal}
\end{align}
where we have defined the functions (for $j=0,1$)
\begin{align}
    \Lambda ^{(\nu )} _{j} (\rho ) =  \int_0^\infty \!\! dk \, \frac{J_{\nu} (k\rho)}{\sqrt{1 + (k l _{0}) ^{2}}} \, e^{-k (2j z_{0} )} . \label{F_Lambda_MainT}
\end{align}

The mathematical details of the derivation of these functions are presented in the Appendix~\ref{ap:scalarpotential}. This result can be interpreted as follows. The first term corresponds to the potential due to the original ionized impurity of charge $Q$ at $z_{0}$. The second term is due to the image of the ionized impurity, of strength $\kappa Q$ localized at the image point $- z_{0}$. The third term $\Lambda ^{(0 )} _{0} (\rho)$ corresponds to the electrostatic potential on the monolayer graphene due to the electronic cloud described by the Yukawa term in Eq. (\ref{SourceYukawa}), and the last term, $\kappa  \Lambda ^{(0 )} _{1} (\rho)$, is the image of such electronic cloud. Figure~\ref{fig:potential_theta_tildes} shows the behavior of the scalar potential of Eq.~\eqref{Eq:Scalarpotentialfinal} with $z_0=1.42\AA$, as a function of $\rho/a_0$, where $a_0=2.46\AA$ is the lattice constant of graphene.

We now evaluate the vector potential ${\bf{A}} ({\bf{r}})$, which is given by Eq. (\ref{VectPotential}) with the vector Green's function of Eq. (\ref{Green_Vector}) and the charge density $\rho _\text{Y} (\mathbf{r})$ of Eq. (\ref{SourceYukawa}). After some manipulations, fully discussed in the Appendix~\ref{ap:scalarpotential}, we get
\begin{align}
    {\bf{A}} ({\boldsymbol{\rho}}) &= {\bf{A}} _\text{Sch} ({\boldsymbol{\rho}}) - \frac{\mu _{1} Q g}{4 \pi } \, \hat{{\bf{e}}}_{\phi} \, \Lambda ^{(1)} _{1} (\rho ) , \label{Eq:VectorPotential}
\end{align}
where the first term ${\bf{A}} _\text{Sch} ({\bf{r}}) $ is exactly the Schwinger's vector potential of a straight vortex line or Dirac string over the $z$ axis
\begin{align}
    {\bf{A}} _\text{Sch} ({\boldsymbol{\rho}}) = \frac{\mu _{1} Q g}{4 \pi } \,   \frac{\hat{{\bf{e}}}_{\phi}}{\rho} \left[ 1 - \frac{2 z _{0}}{\sqrt{\rho ^{2} + (2z _{0}) ^{2}} } \right] ,
\end{align}
which describes a magnetic monopole of strength $Qg$ at the image point $-z_{0}$. The second term, proportional to  $\hat{{\bf{e}}}_{\phi} \, \Lambda ^{(1)} _{1} (\rho )$, corresponds to the magnetic response due to the electronic cloud described by the Yukawa term.

Applying our previous explicit results for the electromagnetic response of the TI, we will study the electrical conductivity in the coupled graphene monolayer. For this purpose, we first need to analyze the scattering mechanism experienced by the massless Dirac fermions in the graphene monolayer, due to the presence of the local electromagnetic fields resulting from this coupling, which is the subject of the next section.

\section{Scattering Analysis and phase shift}
In this section, we shall analyze the scattering mechanism experienced by massless Dirac fermions in the graphene monolayer coupled to the planar TI, as shown in the hetero-structure depicted by Fig.~\ref{fig:scheme_ion}. As discussed in the previous section, the electromagnetic coupling between the TI and the graphene monolayer in the presence of ionized impurities, will generate a local electromagnetic field characterized by the scalar and vector potential $\Phi(\mathbf{r})$ and $\mathbf{A}(\mathbf{r})$, as given by Eq.~\eqref{Eq:Scalarpotentialfinal} and Eq.~\eqref{Eq:VectorPotential}, respectively. Therefore, the dynamics of the charge carriers is determined by the effective Hamiltonian (with the minimal coupling prescription):
\begin{equation}
	\hat{H}^{\xi}=\xi v_\text{F}\bs{\sigma}\cdot \left[ \mbf{\hat{p}}- q\mbf{A}(\mathbf{r})\right]   + q\Phi(\mathbf{r})\hat{I}
	\label{eq:complete_H},
\end{equation}
where $\xi=\pm1$ is the valley index for each of the Dirac $K_{\pm}$ points, $v_\text{F}$ is the Fermi velocity, and $q = \mp e$ is the fermion's electric charge (electrons or holes depending on the sign of the chemical potential). 

As seen in Eq.~\eqref{Eq:Scalarpotentialfinal} and Eq.~\eqref{Eq:VectorPotential}, the electromagnetic fields generated by the coupling decay at long distances with respect to the position of the impurity. Therefore,  we can apply the standard assumptions in scattering theory, i.e. that  incident fermions far from the impurity are described by asymptotically free particle states, with momentum $\mathbf{k}$, and band index $\lambda$. In polar coordinates, those are given by the bi-spinors
\begin{equation}
	 \left\langle \mathbf{x}\middle|\Phi_{\mathbf{k},\lambda}\right\rangle = \Phi_{\mathbf{k}}^{(\lb)}(\mathbf{x})=\frac{1}{\sqrt{2}}\begin{bmatrix}
			1 \\ \lb
		\end{bmatrix}e^{\ii\mathbf{k}\cdot \mathbf{x}}=\frac{1}{\sqrt{2}}\begin{bmatrix}
		1 \\ \lb
	\end{bmatrix}e^{\ii kr\cos \f},\label{eq:incident_spinor}
	\end{equation}
	with energy
	\begin{equation}
	\mathcal{E}^{(0,\xi)}_{\mu,\mathbf{k}}=\lambda\xi\hbar v_\text{F} |\mathbf{k}|. \label{eq:unpert_spectrum}
\end{equation}

Now, it is convenient to use the identity 
\begin{align}
	e^{\ii kr\cos\phi} = \sum_{m=-\infty}^{\infty}\ii^{m} e^{\ii m\phi} J_m(kr),
\end{align}
in order to expand the incident spinor into angular momentum channels $m \in \mathbb{Z}$
\bea
     \Phi_{\mathbf{k}}^{(\lb)}(\mathbf{x})&=&\frac{1}{\sqrt{2}}\sum_{m=-\infty}^{\infty}\ii^m\begin{bmatrix}
		J_{m}(kr)e^{\ii m\f} \\ i\lb J_{m+1}(kr)e^{\ii (m+1)\f}
	\end{bmatrix}.\label{eq:incident_spinor_angular}
\eea

Due to the azymuthal symmetry of the system, we can also expand the angular dependence of the eigenspinor of the full Hamiltonian into angular momentum channels 
	\bea
	\left\langle \mathbf{x}\middle|\Psi_{\mathbf{k},\lambda}\right\rangle  &=&\Psi_{\mathbf{k}}^{(\lb)}(\mathbf{x})\nn\\
	&=&\frac{1}{\sqrt{2}}\sum_{m=-\infty}^{\infty}\ii^m\begin{bmatrix}
		f_{m}(kr)e^{\ii m\f} \\ \ii\lb g_{m}(kr)e^{\ii (m+1)\f}
	\end{bmatrix}\label{eq:full_spinor_angular},
\eea
where the radial dependence is implicit in the functions $f_m(kr)$ and $g_m(kr)$, that are yet to be determined, as we show in Appendix~\ref{sec:GreensFunctionCalculation}. From the Lippmann-Schwinger formalism in terms of the retarded and advanced Green's functions, the final asymptotic states can also be decomposed into angular-momentum channels. To do so, we consider that the electromagnetic potential possesses a compact support, i.e. it decays with a characteristic distance $a$, so that for $r>a$ the interaction potential becomes negligible. Then, the eigenspinors can be found from the following self-consistent integral equations:
\begin{widetext}
{\small
\begin{align}
\begin{bmatrix}
	f_m(r) \\ g_m(r)
\end{bmatrix}&=\begin{bmatrix}
J_m(kr) \\  J_{m+1}(kr)
\end{bmatrix} -\frac{\lb\xi \ii \p  k}{2\hbar v_\text{F}}\int_{0}^{r} dr' r' \begin{bmatrix}
J_m(kr')H_m^{(1)}(kr)& J_{m+1}(kr')H_m^{(1)}(kr) \\
 J_{m}(kr')H_{m+1}^{(1)}(kr) & J_{m+1}(kr')H_{m+1}^{(1)}(kr) 
\end{bmatrix} \begin{bmatrix}
e\Phi(r')& -\lb\xi qv_\text{F} A(r') \\
-\lb\xi qv_\text{F} A(r') & e\Phi(r')
\end{bmatrix}\begin{bmatrix}
f_m(r') \\ g_m(r')
\end{bmatrix}\notag \\
&\quad -\frac{\lb\xi \ii \p  k}{2\hbar v_\text{F}}\int_{r}^{a} dr' r' \begin{bmatrix}
J_m(kr)H_m^{(1)}(kr')&  J_{m+1}(kr)H_m^{(1)}(kr') \\
 J_{m}(kr)H_{m+1}^{(1)}(kr') & J_{m+1}(kr)H_{m+1}^{(1)}(kr') 
\end{bmatrix} \begin{bmatrix}
e\Phi(r')&- \lb \xi qv_\text{F} A(r') \\
- \lb \xi qv_\text{F} A(r') & e\Phi(r')
\end{bmatrix}\begin{bmatrix}
f_m(r') \\ g_m(r')
\end{bmatrix}. \label{eq:integral_eqn_final01}
\end{align}
}
and for $r>a$  we have
{\small
\begin{align}
&\begin{bmatrix}
	f_m(r) \\ g_m(r)
\end{bmatrix}=\begin{bmatrix}
J_m(kr) \\  J_{m+1}(kr)
\end{bmatrix}-\frac{\lb\xi \ii \p  k}{2\hbar v_\text{F}}\int_{0}^{a} dr' r' \begin{bmatrix}
J_m(kr')H_m^{(1)}(kr)&  J_{m+1}(kr')H_m^{(1)}(kr) \\
 J_{m}(kr')H_{m+1}^{(1)}(kr) & J_{m+1}(kr')H_{m+1}^{(1)}(kr) 
\end{bmatrix} \begin{bmatrix}
e\Phi(r')& - \lb \xi ev_\text{F} A(r') \\
- \lb \xi ev_\text{F} A(r') & e\Phi(r')
\end{bmatrix}\begin{bmatrix}
f_m(r') \\ g_m(r')
\end{bmatrix},\label{eq:integral_eqn_final_02}
\end{align}
}
\end{widetext}
which we label as {\it internal} and {\it external} solutions, respectively. 

From the above states, the so-called phase shift $\delta_m$ is easily computed from the {\it external} solution (because it carries the scattering information) as
\begin{widetext}
\bea
e^{\ii\delta_m(k)}\sin \delta_m(k)=-\frac{\lb\xi \p  k}{4\hbar v_\text{F}}\int_{0}^{a} dr' r' \begin{bmatrix}
		J_m(kr') & J_{m+1} (kr')
	\end{bmatrix} \begin{bmatrix}
		q\Phi(r')& -\lb \xi qv_\text{F} A(r') \\
		-\lb \xi qv_\text{F} A(r') & q\Phi(r')
	\end{bmatrix}\begin{bmatrix}
		f_m(r') \\ g_m(r')
	\end{bmatrix},
 \label{eq:phase_shift_final}
\eea
\end{widetext}
so that the spinor $\left[f_m(r'),g_m(r')\right]^T$ is given by the internal solution (because it is the region that produced the scattering).

\section{Relaxation time and Electrical conductivity}
In order to compute the microscopic transport coefficients, we define the quantum operators, related to the particle, energy and heat currents, respectively:
 \begin{equation}
 \hat{\mathbf{j}}=\sum_{\mathbf{p}\sigma}\mathbf{v}_{\mathbf{p}}\hat{n}_{\mathbf{p}\sigma},
 \end{equation}
  \begin{equation}
 \hat{\mathbf{j}}_{E}=\sum_{\mathbf{p}\sigma}\mathbf{v}_{\mathbf{p}}\mathcal{E}_{\mathbf{p}\sigma}\hat{n}_{\mathbf{p}\sigma},
 \end{equation}
\begin{equation}
 \hat{\mathbf{j}}_{Q}=\sum_{\mathbf{p}\sigma}\mathbf{v}_{\mathbf{p}}\left(\mathcal{E}_{\mathbf{p}\sigma}-\mu\right)\hat{n}_{\mathbf{p}\sigma},
 \end{equation}
 where for a particle with momentum $\mathbf{p}$ and spin $\sigma$, we identify $\mathbf{v}_{\mathbf{p}}$ as the group velocity, $\hat{n}_{\mathbf{p}\sigma}$ as the particle number density operator, $\mathcal{E}_{\mathbf{p}\sigma}$ as the energy, and $\mu$ as the chemical potential of the system. The corresponding macroscopic observed currents are given by the ensemble average of the above operators, i.e.,
  \begin{subequations}
 \begin{align}
     \mathbf{J}&=\left\langle  \hat{\mathbf{j}} \right\rangle,\\
     \mathbf{J}_{E}&=\left\langle  \hat{\mathbf{j}}_{E} \right\rangle,\\
     \mathbf{J}_{Q}&=\left\langle  \hat{\mathbf{j}}_{Q} 
     \right\rangle.
 \end{align}
 \end{subequations}

As is shown in Appendix~\ref{sec:The Onsager's coefficients}, the currents are coupled to the temperature and electrochemical potential gradients through the so-called Onsager's coefficients, which in tensor notation take the form\cite{Munoz_2012}:
 \begin{subequations}
     \begin{align}
     \mathbf{J}&=-\frac{1}{T}\overleftrightarrow{\mathbf{L}}^{(11)}\cdot\nabla\left(\mu+eV\right)+\overleftrightarrow{\mathbf{L}}^{(12)}\cdot\nabla\left(\frac{1}{T}\right),\\
     \mathbf{J}_{Q}&=-\frac{1}{T}\overleftrightarrow{\mathbf{L}}^{(21)}\cdot\nabla\left(\mu+eV\right)+\overleftrightarrow{\mathbf{L}}^{(22)}\cdot\nabla\left(\frac{1}{T}\right).
 \end{align}
 \end{subequations}
Here $T$ is the system's temperature and $V$ the external bias voltage that triggers the electric current. Then, the electrical conductivity tensor is found by imposing the isothermal condition  $\nabla T=0$ 
 \begin{equation}
\overleftrightarrow{\boldsymbol{\sigma}}=\frac{e^2}{T}\overleftrightarrow{\mathbf{L}}^{(11)},
 \end{equation}
the thermal conductivity tensor is defined by $\mathbf{J}=0$
 \begin{equation}
     \overleftrightarrow{\boldsymbol{\kappa}}=\frac{1}{T^2}\left(\overleftrightarrow{\mathbf{L}}^{(22)}-\overleftrightarrow{\mathbf{L}}^{(21)}\cdot\left[\overleftrightarrow{\mathbf{L}}^{(11)}\right]^{-1}\cdot\overleftrightarrow{\mathbf{L}}^{(12)} \right),
 \end{equation}
 and the Seebeck coefficient or thermopower is given by
  \begin{equation}
     S=\frac{1}{eT}\left[\overleftrightarrow{\mathbf{L}}^{(11)}\right]^{-1}\cdot\overleftrightarrow{\mathbf{L}}^{(12)}.
 \end{equation}

The connection of the Onsager's coefficients with the microscopic dynamical variables is given by the Kubo's linear response theory. In the Appendix~\ref{The Linear Response Regime} we show that for the particle current operator
\begin{equation}
    \hat{\mathbf{j}}^{(\xi)}(\mathbf{x})=\xi v_\text{F} \left|\mathbf{x}\middle\rangle\boldsymbol{\sigma}\middle\langle\mathbf{x}\right|,
\end{equation}
and the heat current operator
\begin{equation}
    \hat{\mathbf{j}}^{(\xi)}_{Q}(\mathbf{x})=\xi v_\text{F} (\hat{H}^{\xi}-\mu)\left|\mathbf{x}\middle\rangle\boldsymbol{\sigma}\middle\langle\mathbf{x}\right|,
\end{equation}
given the thermal equilibrium density operator 
\begin{equation}
    \hat{\rho}_0=\frac{\exp\left[-\beta \left(\hat{H}^{\xi}-\mu\right)\right]}{\Tr\left[\exp\left[-\beta \left(\hat{H}^{\xi}-\mu\right)\right]\right]},
\end{equation}
the Onsager's coefficients take the form\cite{Mahan}
 \begin{subequations}
    \begin{align}
     L_{\alpha\beta}^{(11)}&=-T\int_{0}^{\infty}dt\,e^{-s\,t}\int_{0}^{\beta}d\beta'\,\Tr \left[\hat{\rho}_{0}\hat{j}_{\alpha}(-t-i\hbar \beta')\hat{j}_{\beta}\right],\\
     L_{\alpha\beta}^{(12)}&=L_{\alpha\beta}^{(21)}\nn\\
     &=-T\int_{0}^{\infty}dt\,e^{-s\,t}\int_{0}^{\beta}d\beta'\,\Tr \left[\hat{\rho}_{0}\hat{j}_{Q\alpha}(-t-i\hbar \beta')\hat{j}_{\beta}\right],\\
     L_{\alpha\beta}^{(22)}&=-T\int_{0}^{\infty}dt\,e^{-s\,t}\int_{0}^{\beta}d\beta'\,\Tr \left[\hat{\rho}_{0}\hat{j}_{Q\alpha}(-t-i\hbar \beta')\hat{j}_{Q\beta}\right].
 \end{align} 
 \end{subequations}

 These expressions are reduced to a more explicit form by introducing the spectral density function \cite{nano12203711}
\begin{eqnarray}
&&\boldsymbol{\mathcal{A}}^{\xi}(\mathbf{x},\mathbf{x'};E)\nn\\
&=& 2\pi\sum_{\lambda}\int\frac{d^2k_{\parallel}}{(2\pi)^2}
\Psi_{\lambda,\mathbf{k}_{\parallel}}(\mathbf{x})\otimes \Psi_{\lambda,\mathbf{k}_{\parallel}}^{\dagger}(\mathbf{x'} )\delta\left(E-\mathcal{E}^{\lambda,\xi}_{\mathbf{k}_{\parallel}} \right),\nn\\
\end{eqnarray}
and its relationship with the retarded and advanced Green's functions:
\begin{eqnarray}
\mathcal{A}^{\lambda,\xi}(k_{\parallel};E)
= \ii \left[ \langle G_{R}^{\lambda,\xi}(k_{\parallel};E)  \rangle - \langle G_{A}^{\lambda,\xi}(k_{\parallel};E)  \rangle\right],
\end{eqnarray}
so that, for instance, the coefficient $L_{\alpha\beta}^{(11)}$ takes the form:
\begin{eqnarray}
L_{\alpha\beta}^{(11)}(T) &=& \delta_{\alpha\beta}4\pi\left( \frac{\hbar v_\text{F}^2 T}{\left(2\pi\right)^3} \right) \int_{-\infty}^{\infty}dE \left( -\frac{\partial f_0 (E)}{\partial E} \right)\nn\\
&\times&\int_{0}^{\infty} dk_{\parallel} \langle G_{R}^{\lambda,\xi}(k_{\parallel};E)  \rangle \langle G_{A}^{\lambda,\xi}(k_{\parallel};E)  \rangle \frac{\mathbf{k}_{\parallel}\cdot\mathbf{k}_{\parallel}}{k_{\parallel}}.\nonumber\\
\label{eq_L_uncorrected}
\end{eqnarray}
where $f_0(E,T) = \left(1 + \exp[(E-\mu)/kT]\right)^{-1}$ is the Fermi-Dirac distribution.

\begin{figure}[h!]
		\centering
		\includegraphics[width=1 \columnwidth]{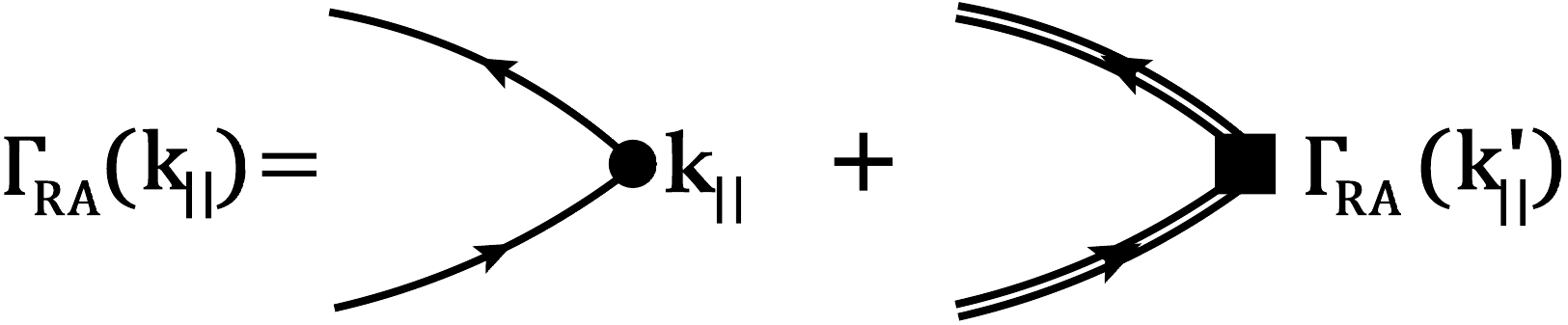}
		\caption{The Bethe-Salpeter integral equation for the vertex function $\boldsymbol{\Gamma}_\text{RA}(\mathbf{k})$.}
		\label{fig:bethe_salpeter}
	\end{figure}

By following Ref.~\cite{nano12203711} (see Appendix~\ref{sec:The Vertex Corrections and Relaxation Time} for details) it is possible to include vertex corrections to the formalism. For that purpose, one of the $\mathbf{k}_{\parallel}$ factors is replaced by the vertex correction $\boldsymbol{\Gamma}_\text{RA}(\mathbf{k}_{\parallel},E)$, that satisfies the {\it Bethe-Salpeter equation} (see Fig.~\ref{fig:bethe_salpeter}):
{\small
  \begin{align}
		&\boldsymbol{\Gamma}_\text{RA}(\mathbf{k}_{\parallel},E)=\mathbf{k}_{\parallel}\nn\\
  &+n_\text{imp}\int \frac{d^2k'_{\parallel}}{(2\pi)^2} \left\langle G_R^{\lambda,\xi}(\mathbf{k'}_{\parallel})\right\rangle\left\langle G_A^{\lambda,\xi}(\mathbf{k'}_{\parallel})\right\rangle\left| \hat{T}^{(\lambda,\xi)}_{\mathbf{k'}_{\parallel}\mathbf{k}_{\parallel}} \right|^2\boldsymbol{\Gamma}_\text{RA}(\mathbf{k'}_{\parallel},E),
	    \end{align}
}
where $\hat{T}^{(\lambda,\xi)}_{\mathbf{k'}_{\parallel}\mathbf{k}_{\parallel}}$ is the $T$-matrix operator. 

 Therefore, when vertex corrections are incorporated then Eq.~\eqref{eq_L_uncorrected} (as well as the other coefficients) is modified as
 \begin{align} 
	   & L^{(11)}_{\alpha\beta}(T)=\delta_{\alpha\beta}\frac{\hbar v_{F}^2 T}{2\pi^2 } \int_{-\infty}^{\infty}dE\,\,\left(-\frac{\partial f_0(E)}{\partial E}\right)\nn\\
     &\times\int_{0}^{\infty} dk_{\parallel} \left\langle G_R^{\lambda,\xi}(\mathbf{k}_{\parallel})\right\rangle\left\langle G_A^{\lambda,\xi}(\mathbf{k}_{\parallel})\right\rangle\frac{\mathbf{k}_{\parallel}\cdot\boldsymbol{\Gamma}_\text{RA}(\mathbf{k}_{\parallel},E)}{k_{\parallel}} , 
	    \end{align}

     \begin{align}
	   &L^{(12)}_{\alpha\beta}(T)\notag\\&=\delta_{\alpha\beta}\frac{\hbar v_{F}^2 T}{2\pi^2 } \int_{-\infty}^{\infty}dE\,\,\left(-\frac{\partial f_0(E)}{\partial E}\right)(E - \mu)\nn\\ &\times\int_{0}^{\infty} dk_{\parallel} \left\langle G_R^{\lambda,\xi}(\mathbf{k}_{\parallel})\right\rangle\left\langle G_A^{\lambda,\xi}(\mathbf{k}_{\parallel})\right\rangle\frac{\mathbf{k}_{\parallel}\cdot\boldsymbol{\Gamma}_\text{RA}(\mathbf{k}_{\parallel},E)}{k_{\parallel}}
     ,
	    \end{align}
     and 
\begin{align}
	    &L^{(22)}_{\alpha\beta}(T)\notag\\&=\delta_{\alpha\beta}\frac{\hbar v_{F}^2 T}{2\pi^2 } \int_{-\infty}^{\infty}dE\,\,\left(-\frac{\partial f_0(E)}{\partial E}\right)(E - \mu)^2\nn\\ &\times\int_{0}^{\infty} dk_{\parallel} \left\langle G_R^{\lambda,\xi}(\mathbf{k}_{\parallel})\right\rangle\left\langle G_A^{\lambda,\xi}(\mathbf{k}_{\parallel})\right\rangle\frac{\mathbf{k}_{\parallel}\cdot\boldsymbol{\Gamma}_\text{RA}(\mathbf{k}_{\parallel},E)}{k_{\parallel}}. 
	    \end{align}

Following Ref.~\cite{nano12203711}, as is shown in Appendix~\ref{sec:The Vertex Corrections and Relaxation Time}, by considering the general form $\boldsymbol{\Gamma}_\text{RA}(\mathbf{k}_{\parallel},E)=\gamma(\mathbf{k}_{\parallel},E)\mathbf{k}_{\parallel}$, the relaxation time can be introduced with the relation
\begin{equation}
	    \gamma(k_\text{F})=\frac{\tau_1(k_\text{F})}{\tau_1(k_\text{F})-\tau(k_\text{F})},
\end{equation}
 where we defined (for $\cos\phi'=\mathbf{k}_{\parallel}\cdot\mathbf{k'}_{\parallel}/k^2_{\parallel}$)
\bea
	    &&\frac{1}{\tau_1(k_\text{F})}\nn\\
     &=&\frac{2\pi n_\text{imp}}{\hbar}\int \frac{d^2k'_{\parallel}}{(2\pi)^2}\left| T^{(\lambda,\xi)}_{\mathbf{k'}_{\parallel}\mathbf{k}_{\parallel}} \right|^2\cos \phi'\,\,\delta(\hbar v_\text{F} k_\text{F}-\hbar v_\text{F} k'_{\parallel}).\nn\\
\eea
so that, by following Ref.~\cite{nano12203711} the total \textit{transport relaxation time} is defined by
	\begin{align}
	    &\frac{1}{\tau_{\text{tr}}(k_\text{F})}=\frac{1}{\tau(k_\text{F})}-\frac{1}{\tau_1(k_\text{F})}\\
	    &=\frac{2\pi n_\text{imp}}{\hbar}\int \frac{d^2k'}{(2\pi)^2}\delta(\hbar v_\text{F} k_\text{F}-\hbar v_\text{F} k')\left| T^{(\lambda,\xi)}_{\mathbf{k'}_{\parallel}\mathbf{k}_{\parallel}} \right|^2(1-\cos \phi'),\nonumber
	\end{align}
which can be expressed in terms of the scattering phase shifts $\delta_m(k)$ of Eq.~\eqref{eq:phase_shift_final} as
\begin{equation}
	    \frac{1}{\tau_{\text{tr}}(k_\text{F})}=\frac{2 n_\text{imp}  v_\text{F}}{k_\text{F}}\sum_{m=-\infty}^{\infty}\sin^2 \left[\delta_{m}(k_\text{F})-\delta_{m-1}(k_\text{F}) \right].
     \label{eq_tautr}
\end{equation}

With all these ingredients, as is computed in Appendixes~\ref{sec:The Electrical Conductivity} and~\ref{sec:The Thermal Conductivity and Seebeck Coefficient}, the electrical conductivity is
 \bea
\sigma_{xx}(T)&=&4\left(\frac{e^2}{h}\right)k_\text{F}v_\text{F}~\tau_\text{tr}(k_\text{F})\nn\\
&\times&\left[1+2\frac{k_\text{B}T}{v_\text{F}\hbar k_\text{F}}\ln\left(1+\exp\left[-\frac{\hbar k_\text{F} v_\text{F}}{k_\text{B}T}\right]\right)\right],\nn\\
\label{eq:sigmaxx}
\eea
and the thermal conductivity and the Seebeck coefficient are, respectively, given by
\begin{align}
    &\kappa_{\alpha\alpha}(T)=-\frac{2\hbar^2}{\pi k_\text{B} T^2}\left(\frac{k_\text{B} T}{\hbar }\right)^4\tau_{\text{tr}}(k_\text{F})\nn\\
    &\times\sum_{\lambda,\xi=\pm1}\left[3\,\text{Li}_3\left(-e^{\frac{\lambda\xi\hbar v_\text{F} k_\text{F}}{k_\text{B} T}}\right)+2\frac{\left[\text{Li}_2\left(-e^{\frac{\lambda\xi\hbar v_\text{F} k_\text{F}}{k_\text{B} T}}\right)\right]^2}{\ln\left(1 + e^{\frac{\lambda\xi\hbar v_\text{F} k_\text{F}}{k_\text{B} T}} \right)}\right].
\end{align}
\begin{align}
  S(T)&=\frac{1}{eT}\sum_{\lambda,\xi}\frac{L^{(12)(\lambda,\xi)}_{\alpha\alpha}(T)}{L^{(11)(\lambda,\xi)}_{\alpha\alpha}(T)}\nn\\
  &=-\frac{k_\text{B}}{e}\sum_{\lambda=\pm1}\sum_{\xi=\pm1}\left( \frac{2\lambda\xi\text{Li}_2\left(-e^{\frac{\lambda\xi\hbar v_\text{F} k_\text{F}}{k_\text{B} T}}\right)}{\ln\left(1 + e^{\frac{\lambda\xi\hbar v_\text{F} k_\text{F}}{k_\text{B} T}} \right)}+\frac{\hbar v_\text{F} k_\text{F}}{k_\text{B} T}\right).
\end{align}

\section{Results and discussion}\label{Sec:Results and discussion}
In the following, we set the Fermi velocity in graphene as $v_\text{F}=10^{16}\AA s^{-1}$, the chiral and band indexes as $\xi=1$, and $\lambda=1$, and the distance between the TI surface and the graphene monolayer as $z_0=1.42\AA$. As a function of the free carrier density in grapgene $n_c$, we obtain the Fermi $k_\text{F}$, and the Thomas-Fermi $q_\text{TF}$ wave-vectors { as defined by Eq.~\eqref{eq:kF} and Eq.~\eqref{eq:qTF2}, respectively, and the corresponding Yukawa screening length $l_0 = q_\text{TF}^{-1}$.}
%
%
Moreover, {for a monovalent ionic impurity the charge $Q = +e$}, all the magnetic permeabilities are fixed to unity, and the relative dielectric permittivity $\epsilon_2$ is taken as $6.9$, corresponding to graphene for which we assume $\theta_2=0$.

\subsection{The role of the MEP $\theta_1$}\label{sec:The impact of theta}
To test the impact of the topological component, represented by the MEP parameter $\theta_1$, on the electromagnetic coupling between the TI and the massless Dirac fermions in graphene, we compute the electrical conductivity from Eq.~\eqref{eq:sigmaxx} for two different materials: TlBiSe$_2$, and TbPO$_4$. Those materials are characterized by the parameters displayed in Table~\ref{Tab:TIwithThetaExp}, where $\Tilde{\theta}\equiv\alpha\theta_1/\pi$.

\begin{table}[h!]
\begin{tabular}{|P{0.1\textwidth}|P{0.1\textwidth}|P{0.1\textwidth}|P{0.1\textwidth}|}
\hline
\textbf{Material} & $\boldsymbol{\epsilon}_r$ & $\boldsymbol{\tilde{\theta}}$ & \textbf{Ref.} \\ \hline
TlBiSe$_2$        &                   4             &                $11\alpha$                    &        \cite{PhysRevB.105.155120}        \\ \hline
TbPO$_4$          &              3.5                  &              0.22                      &       \cite{franca2021radiation}        \\ \hline
\end{tabular}
\caption{Material constants for TlBiSe$_2$, and TbPO$_4$.}
\label{Tab:TIwithThetaExp}
\end{table}

\begin{figure}[h!]
    \centering
    \includegraphics[scale=0.72]{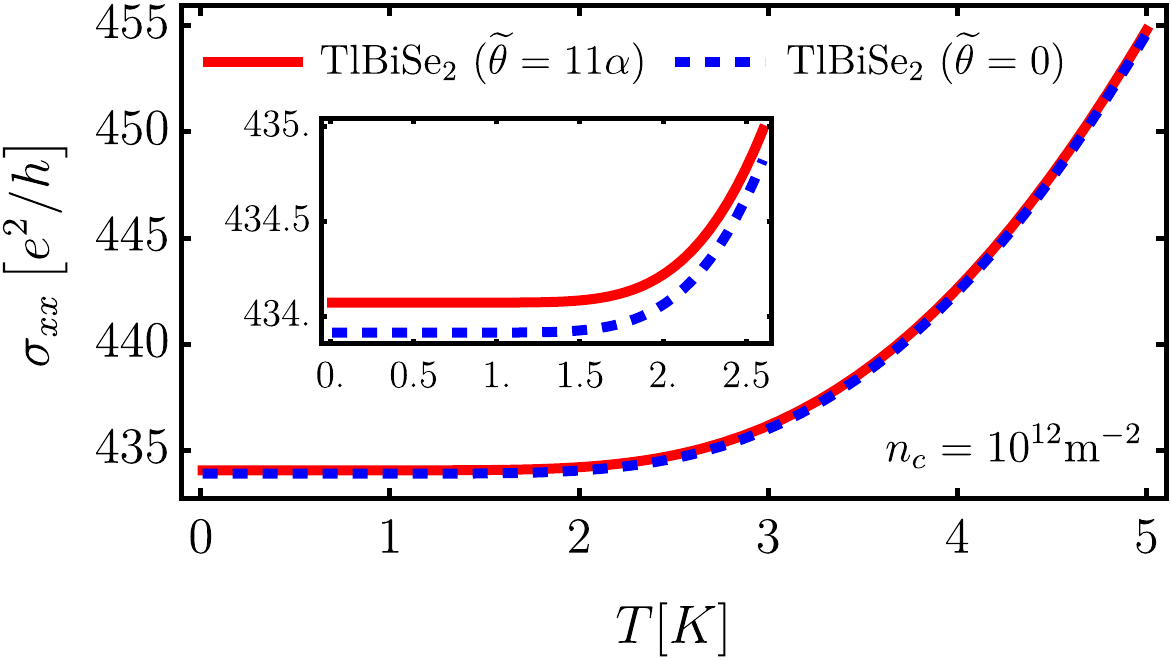}
    \caption{Electrical conductivity in the graphene monolayer $\sigma_{xx}$ as a function of temperature. Each curve corresponds to a TlBiSe$_2$ TI slab, with and without the contribution from the MEP term. The inset is included to appreciate the small deviations at low-temperatures. The impurity concentration is taken as $n_\text{imp}=10^{12}$m$^{-2}$.}
    \label{fig:sigmaxx_TlBiSe2_1012}
\end{figure}

Figure~\ref{fig:sigmaxx_TlBiSe2_1012} shows the electrical conductivity in the graphene monolayer (with a carrier density $n_c=10^{12}\text{m}^{-2}$) computed from Eq.~\eqref{eq:sigmaxx} when the TI slab is made of TlBiSe$_2$. To elucidate the impact of the topological MEP terms, we implemented the cases $\widetilde{\theta}=0$ and $\widetilde{\theta}=11\alpha$, as Table~\ref{Tab:TIwithThetaExp} indicates. As can be noticed, the topological effects are negligible even for low temperatures, where the residual conductivity is modified just by a factor $\sim0.01\%$.

\begin{figure}[h!]
    \centering
    \includegraphics[scale=0.72]{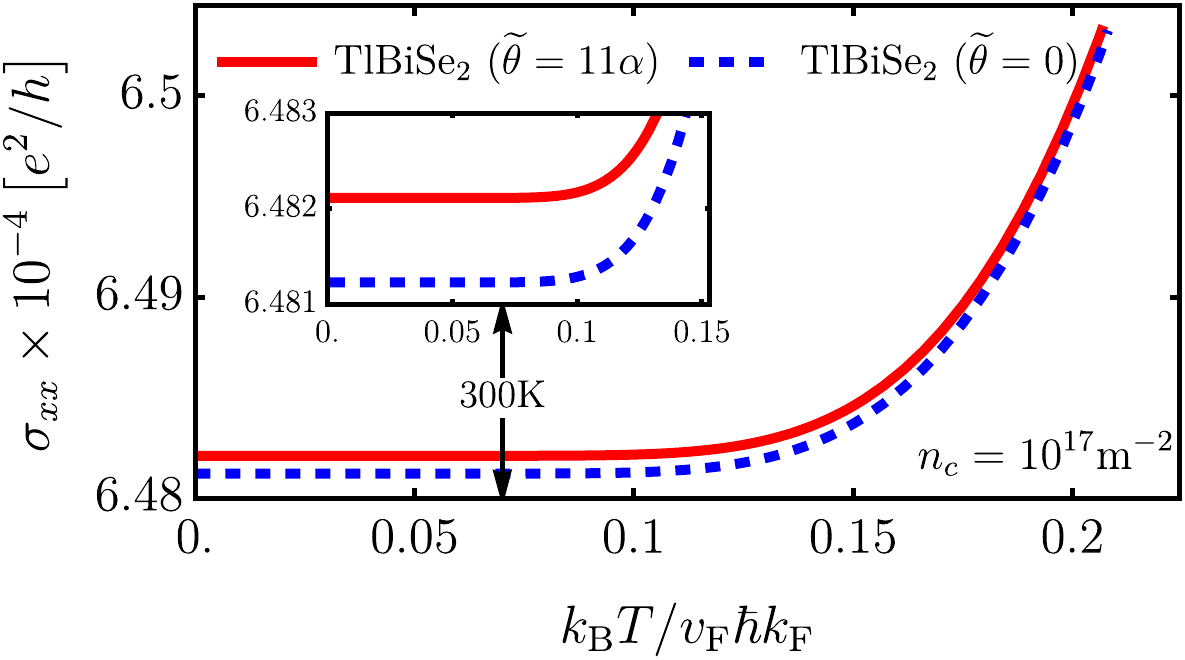}
    \caption{Electrical conductivity in the graphene monolayer $\sigma_{xx}$ (scaled by a factor of $10^{-4}$) as a function of the dimensionless temperature  $k_\text{B}T/v_\text{F}\hbar k_\text{F}$. Each curve corresponds to a slab made of TlBiSe$_2$ with and without the contribution from the MEP. The inset is included to appreciate the small deviations at low-temperatures, and the arrows indicate the absolute temperature $T=300$K. The impurity concentration is taken as $n_\text{imp}=10^{12}$m$^{-2}$.}
    \label{fig:sigmaxx_TlBiSe2}
\end{figure}

When the carrier density is increased to $n_c=10^{17}\text{m}^{-2}$, the picture doesn't change: as is depicted in Fig.~\ref{fig:sigmaxx_TlBiSe2}, the effect of the topological MEP term remains small. Nevertheless, the electrical conductivity remains essentially constant for temperatures up to $300$K, and therefore the (small) effects of the topological MEP terms remain present at room temperature. The same behavior is found for TbPO$_4$, i.e., the effects of the topological MEP $\theta$-terms on the electrical conductivity remain very small, on the order of $0.1\%$. 

\begin{figure}[h!]
    \centering
    \includegraphics[scale=0.72]{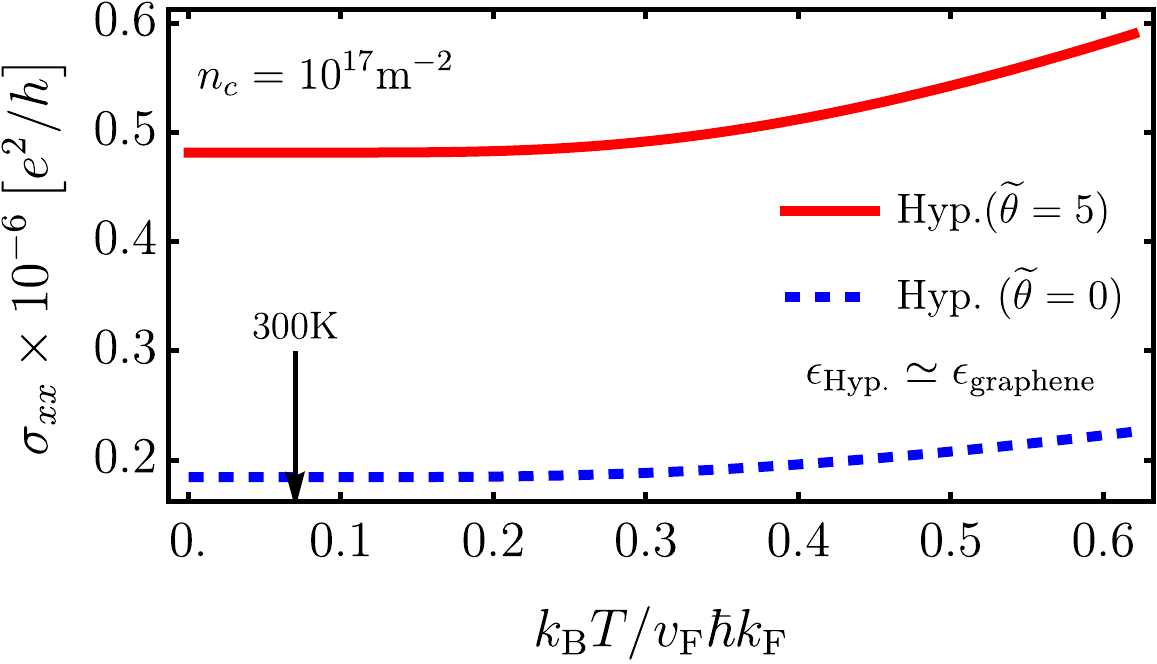}
    \caption{Graphene's electrical conductivity $\sigma_{xx}$ (scaled by a factor of $10^{-6}$) as a function of the dimensionless temperature  $k_\text{B}T/v_\text{F}\hbar k_\text{F}$. Each curve corresponds to a slab made of an hypothetical material with $\epsilon_\text{Hyp.}\simeq\epsilon_\text{graphene}$. The arrow indicates the absolute temperature $T=300$K. The impurity concentration is taken as $n_\text{imp}=10^{12}$m$^{-2}$.}
    \label{fig:sigmaxx_hyp}
\end{figure}

In order to understand the effects of the MEP terms, let us analyze Fig.~\ref{fig:potential_theta_tildes} where the panels are constructed for several values of the Thomas-Fermi wavevector $q_\text{TF}$ and the MEP $\widetilde{\theta}$. As can be appreciated, the topological effects are sensitive to the value of $q_\text{TF}$, which implies that the integration regions for the Eqs.~\eqref{eq:integral_eqn_final01}-\eqref{eq:phase_shift_final} need to be carefully fixed for each window of parameters. For instance, Fig.~\ref{fig:potential_theta_tildes} implies that the scalar potential is negligible for $\rho\gtrsim40a_0$. To have an insight into these results in terms of the intrinsic graphene parameters, the Thomas-Fermi wavevector $q_\text{TF}$ can be calculated as a function of the corresponding free carrier density $n_c$. Therefore, by following Eq.~\eqref{eq:qTF2} we choose the four cases
\bea
n_c^\text{(a)} = 3.2\cdot 10^{13}\text{m}^{-2}&,&~~n_c^\text{(b)} = 3.2\times10^{15}\text{m}^{-2}\nn\\
n_c^\text{(c)} = 3.2\times10^{17} \text{m}^{-2}&,&~~
n_c^\text{(d)} = 3.2\times10^{19} \text{m}^{-2},
\eea
where the super-indexes (a)-(d) correspond to the panels in Figs.~\ref{fig:potential_theta_tildes}~(a)-(d), respectively. Note that despite the hierarchy and deviations between the magnitude of the potential as a function of $\Tilde{\theta}$ remain, for higher carrier densities the topological effects contribute to a slower decay of the potential, that then sustains finite values at longer distances (see Fig.~\ref{fig:potential_theta_tildes}-(d)). Moreover, as seen in the same figure the potential develops a second maxima, that can be attributed to the contribution arising from the MEP topological term. This effect implies that the topological terms are comparatively less screened than the trivial ones. Nevertheless, it is important to point out that the state-of-the-art estimations for the value of $\tilde{\theta}$ suggest that it is small (see Table~\ref{Tab:TIwithThetaExp}). Hence, as Fig.~\ref{fig:potential_theta_tildes} shows, the topological contribution to the electromagnetic coupling can be neglected unless the TI material satisfies $\theta>137\pi$. The later scenario is shown Fig.~\ref{fig:sigmaxx_hyp} where we plot the electrical conductivity for an hypothetical material with $\widetilde{\theta}=3$ and a dielectric constant close to graphene. In this case, even at room temperature the topological effects are considerable so that the electrical conduction in the sample is enhanced. Additionally, as in the case of realistic MEP's values, if the graphene carrier density is increased, the topological effects can be differentiated from the trivial ones up to room temperature. In contrast, for low $n_c$, the topological MEP terms induce appreciable deviations only at low temperatures.  

\subsection{The role of $\epsilon_1$}
An essential characteristic of the scalar potential of Eq.~\eqref{Eq:Scalarpotentialfinal} is its dependence on the dielectric constant  through the $\kappa$ factor in Eq.~\eqref{kappa_and_g}. In particular, by ignoring the topological parameters $\theta$, the combination $(\epsilon_2-\epsilon_1)/(\epsilon_2+\epsilon_1)$ can be understood as an effective image charge located at the slab. Therefore, we present the change in the conductivity as a function of this image charge when the relative permittivity $\epsilon_2$ is fixed as the graphene one. 

\begin{table}[h!]
\begin{tabular}{|P{0.1\textwidth}|P{0.1\textwidth}|P{0.1\textwidth}|}
\hline
\textbf{Material} & $\boldsymbol{\epsilon_r}$ & \textbf{Ref.} \\ \hline\hline
PbTe              & 414                 &       \cite{gibbs2013optical,C1EE01314A}        \\ \hline
Bi$_2$Te$_3$      & 290                 &     \cite{GREENAWAY19651585,https://doi.org/10.1002/aelm.201800904}  \\ \hline
PbSe              & 210                 &    \cite{C3TA14929C,doi:10.1073/pnas.1111419109}           \\ \hline
PbS               & 169                 &      \cite{wang2013material,PhysRevB.35.4511}         \\ \hline
Bi$_2$Se$_3$      & 113                 &     \cite{xia2009observation,https://doi.org/10.1002/aelm.201800904}          \\  \hline
\end{tabular}
\caption{Relative dielectric permittivity $\epsilon_r$ for several topological insulators.}
\label{Tab:TablaEpsilon}
\end{table}

\begin{figure}[h!]
    \centering
    \includegraphics[scale=0.72]{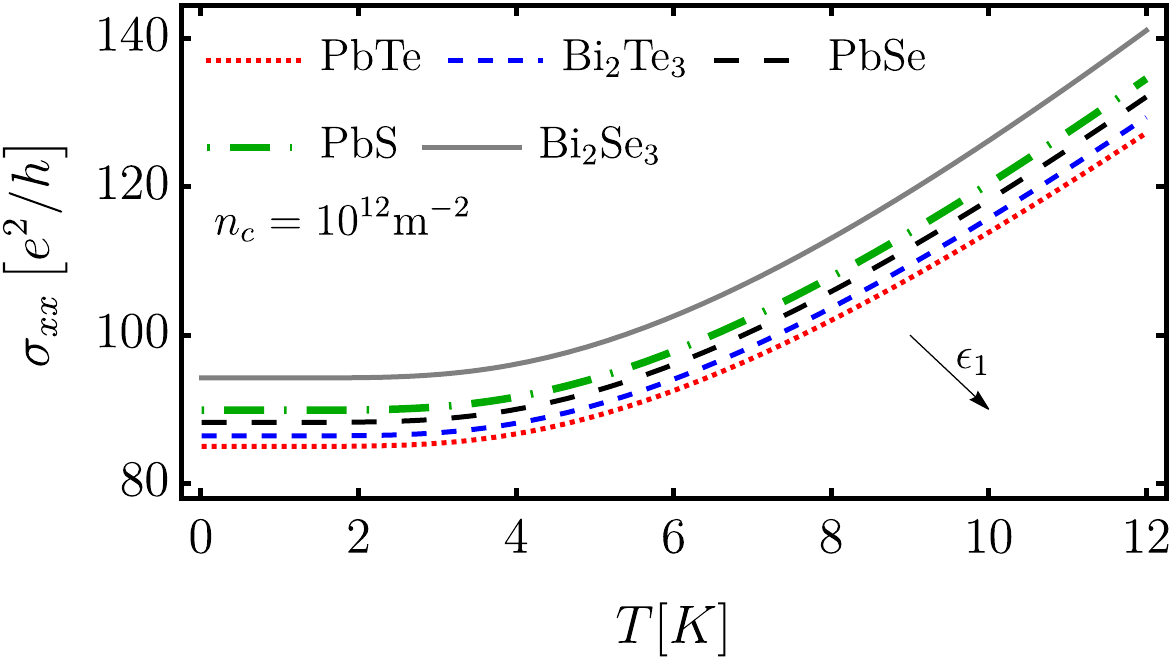}
    \caption{Electrical conductivity of graphene $\sigma_{xx}$ as a function of temperature for $n_c=10^{12}\text{m}^{-2}$. Each curve corresponds to different materials for the slab. The relative permittivity $\epsilon_1$ of each TI is shown in Table~\ref{Tab:TablaEpsilon}. The arrow points the direction in which $\epsilon_1$ increases. The impurity concentration is taken as $n_\text{imp}=10^{12}$m$^{-2}$.}
    \label{fig:sigmaxx_TI}
\end{figure}

Figure~\ref{fig:sigmaxx_TI} shows the electrical conductivity $\sigma_{xx}(k=k_\text{F})$ computed from Eq.~\eqref{eq:sigmaxx}, as a function of temperature for graphene when it is close to different TI materials. Here, due to the results of Sec.~\ref{sec:The impact of theta}, we ignore the MEP $\theta$. In order to reproduce the curves for actual materials, we follow the data of Table~\ref{Tab:TablaEpsilon} where the relative permittivities of the TIs are presented. As can be noticed, at higher TI permittivities the electrical conductivity decreases, consistent with the presence of a strong image charge that considerably modifies the free carrier mobility in graphene.

\begin{table}[h!]
\begin{tabular}{|P{0.1\textwidth}|P{0.1\textwidth}|P{0.1\textwidth}|}
\hline
\textbf{Material} & $\boldsymbol{\epsilon_r}$ & \textbf{Ref.} \\ \hline\hline
Si                & 11.9                &    \cite{madelung2002group,PhysRev.80.72}           \\ \hline
GaAs              & 13.18               &        \cite{strzalkowski1976dielectric,Akinlami_2013}       \\ \hline
InSb              & 17.6                &  \cite{PhysRevB.3.3287,https://doi.org/10.1002/pssb.2220820103}             \\ \hline
\end{tabular}
\caption{Relative dielectric permittivity $\epsilon_r$ for several semiconductors.}
\label{Tab:TablaEpsilonSC}
\end{table}

\begin{figure}[h!]
    \centering
    \includegraphics[scale=0.72]{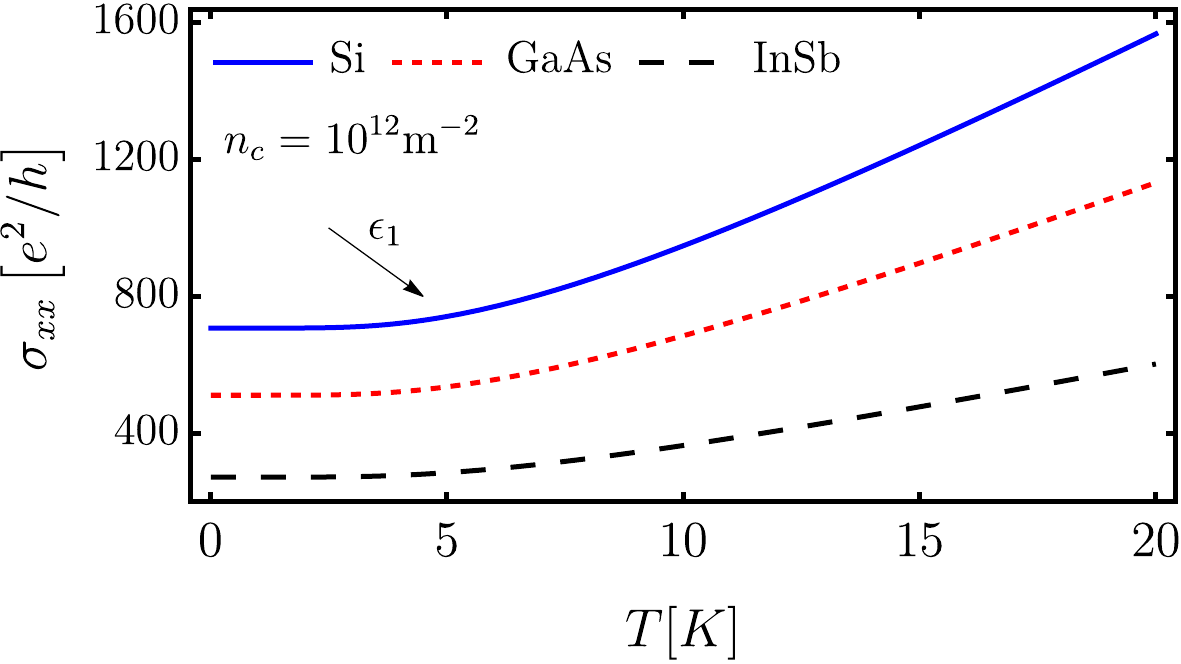}
    \caption{Electrical conductivity of graphene as a function of temperature. Each curve corresponds to different materials for the slab. The relative permittivity of the materials is shown in Table~\ref{Tab:TablaEpsilon}. The arrow points the direction in which $\epsilon_1$ increases. The impurity concentration is taken as $n_\text{imp}=10^{12}$m$^{-2}$.}
    \label{fig:sigmaxx_SC}
\end{figure}

\begin{figure}[h!]
    \centering
    \includegraphics[scale=0.72]{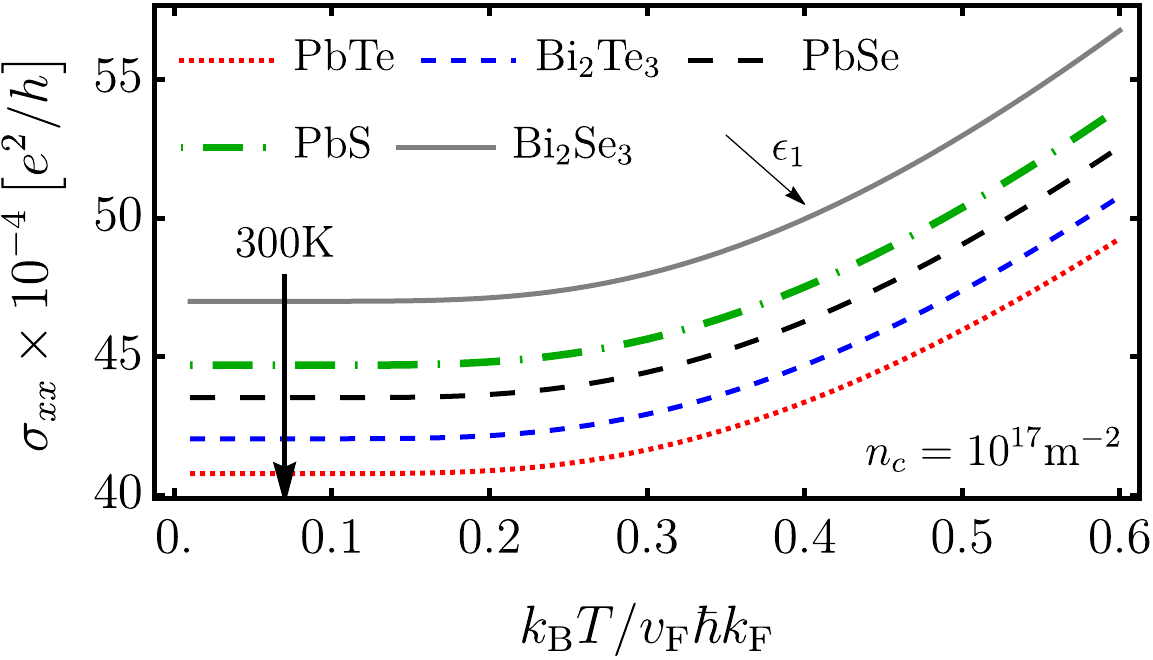}
    \caption{Electrical conductivity of graphene $\sigma_{xx}$ (scaled by a factor of $10^{-4}$) as a function of the dimensionless temperature  $k_\text{B}T/v_\text{F}\hbar k_\text{F}$ for $n_c=10^{17}\text{m}^{-2}$. Each curve corresponds to different materials for the slab. The relative permittivity $\epsilon_1$ of each TI is shown in Table~\ref{Tab:TablaEpsilon}. The thick and thin arrows indicate the absolute temperature $T=300$K and the direction in which $\epsilon_1$ increases, respectively. The impurity concentration is taken as $n_\text{imp}=10^{12}$m$^{-2}$. }
    \label{fig:sigmaxx_TI_1017}
\end{figure}

\begin{figure}[h!]
    \centering
    \includegraphics[scale=0.72]{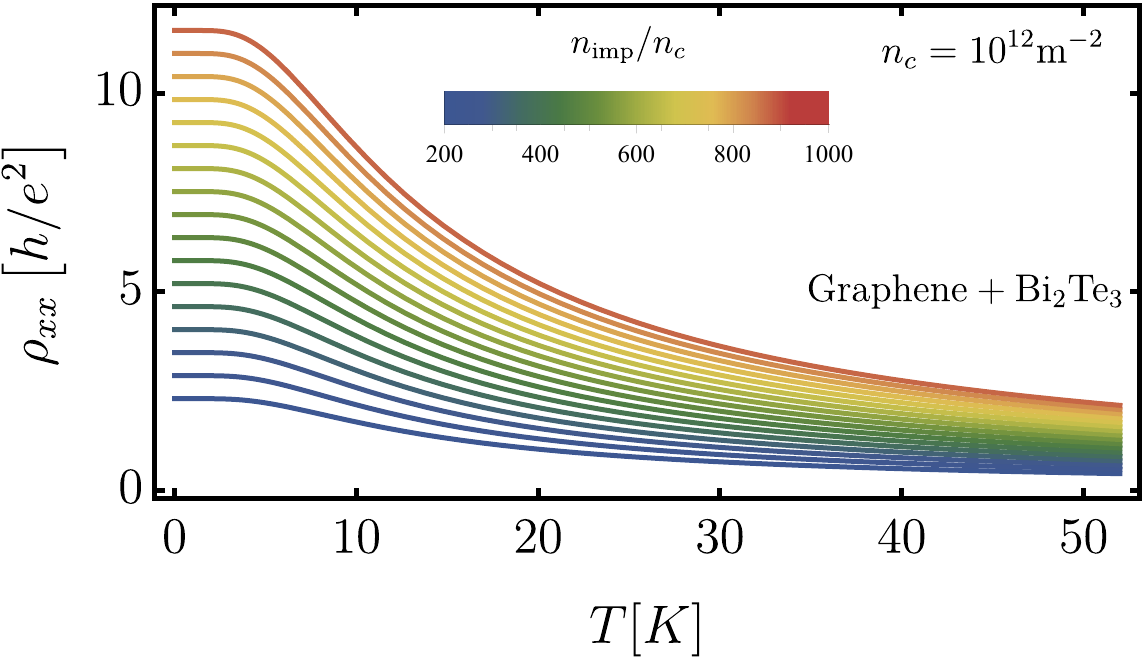}
    \caption{Electrical resistivity of graphene as a function of temperature and the ionized impurity concentration $n_\text{imp} $. Here is assumed that graphene is close to a Bi$_2$Te$_3$ slab.}
    \label{fig:sigmaxx_concentration}
\end{figure}

A similar situation is found when the slab material is changed to a normal semiconductor (Si, GaAs, and InSb), as is shown in Fig.~\ref{fig:sigmaxx_SC}. Such behavior supports the idea of the screening effect due to the image charge on the scattered fermions. In fact, note that for the selected semiconductors, the graphene's conductivity is high compared with the case when the slab is made of a TI. This is because the former materials possess small relative permittivities (see Table~\ref{Tab:TablaEpsilonSC}) in contrast with the latter ones that have $\epsilon_r$ of one order of magnitude higher. 

As may be expected, the carrier density in graphene also modifies its electrical conductivity. For instance, Fig.~\ref{fig:sigmaxx_TI_1017} shows $\sigma_{xx}$ for a carrier concentration of $n_c=10^{17}$m$^{-2}$ when the slab is build of different TIs. Again, the screening effects on the image charge are presented in the scattering process so that the conductivity decreases for higher $\epsilon_r$. Nevertheless, the electrical conductivity remains constant up to room temperature, a feature that may be used for several applications. 

Finally,  Fig.~\ref{fig:sigmaxx_concentration} displays the electrical resistivity $\rho_{xx}\equiv\sigma_{xx}^{-1}$ for graphene when the slab is made up of Bi$_2$Te$_3$, as a function of the impurity concentration $n_\text{imp}$. Clearly, the conductivity increases when there are fewer ions inserted in graphene, given that the fermions have a larger free mean path, as can be directly appreciated from Eq.~\eqref{eq_tautr} where the transport relaxation time is clearly inversely proportional to the impurity concentration $n_\text{imp}$.

\section{Summary and conclusions}\label{Sec:Summary and conclusions}

In this work, we studied the electromagnetic coupling in a hetero-structure made of a TI slab in contact with a graphene monolayer, the later with a diluted concentration of ionized impurities. By taking into account the topologically non-trivial response due to the presence of the MEP terms in the TI, we studied the configuration of the local electromagnetic fields, including the effects of the image charges and the electronic screening due to the presence of the ionized impurities in graphene. As a probe for this electromagnetic coupling, we used Kubo's linear response formalism to calculate the electrical conductivity at finite temperature in graphene under this configuration. To test this theory in a reallistic scenario, we evaluated our analytical formulas for the characteristic parameters of several TI materials. Our result suggest that the contribution to the conductivity arising from the topologically non-trivial MEP terms are in general small, except at very low temperatures and large carrier densities in the graphene monolayer. On the other hand, it was also shown that the difference between the dielectric constants, expressed by the combination $\left( \epsilon_2 - \epsilon_1  \right)/\left( \epsilon_2 + \epsilon_1 \right)$ possesses an important role at determining the overall magnitude of  the electrical conductivity, as can be inferred from our analytical theory since it represents the magnitude of an effective image charge located at the TI slab. We tested this effect by considering the realistic parameters for several different TI materials (Table~\ref{Tab:TablaEpsilon}), including also for comparison the coupling with a few common semiconductors Si, GaAs and InSb (see Fig.~\ref{fig:sigmaxx_SC}). In all those cases, the important effect of the relative dielectric permittivity of the slab material on the electrical conductivity of the coupled graphene monolayer is verified. This rather robust effect suggest a mechanism to control and modulate the electronic transport properties in this type of hetero-structures.


\acknowledgements{ J.D.C.-Y. and E.M. acknowledge financial support from ANID PIA Anillo ACT/192023. E.M. also acknowledges financial support from Fondecyt 1190361. Daniel Bonilla was funded by the ANID Beca Doctorado Nacional 2018 grant No 21180547. A.M.-R. has been partially supported by DGAPA-UNAM Project No. IA102722 and by Project CONACyT (M\'{e}xico) No. 428214}

\appendix
\section{Calculation of the electromagnetic  potentials}\label{ap:scalarpotential}

In Section~\ref{EM_Resp_TI} we presented a detailed derivation of the electromagnetic response of the TI material, as expressed by Eq.~(\ref{Eq:Scalarpotentialfinal}) for the scalar potential and Eq.~(\ref{Eq:VectorPotential}) for the vector potential, respectively.

The scalar potential is given by Eq.~(\ref{ElectricPot}) with the charge density of Eq.~(\ref{SourceYukawa}), i.e. $\Phi ({\bf{r}}) = \int \mathcal{G} ({\bf{r}} , {\bf{r}}^{\prime}) \, \rho _\text{Y}({\bf{r}}^{\prime}) \, d^{3} {\bf{r}}^{\prime}$. After performing the integration over the Dirac's deltas associated with the localized ionized impurity and the screening cloud at the graphene monolayer (located at $z = z_0$), we obtain
\begin{align}
    \Phi (\boldsymbol{\rho}) &= \frac{Q}{4 \pi \epsilon _{1}} \left( \frac{1}{\rho} + \frac{\kappa}{\sqrt{\rho ^{2} + (2 z_{0})^{2} }}\right)\nn\\
    &- \frac{1}{4 \pi \epsilon _{1}} \frac{Q}{2 \pi l_0}  \int \left( \frac{1}{\sqrt{(x-x^{\prime})^{2} + (y-y^{\prime})^{2}}}\right.\nn\\
    &+ \left.\frac{\kappa}{\sqrt{(x-x^{\prime})^{2} + (y-y^{\prime})^{2} + (2z_{0})^{2} }}\right)  \frac{e^{- \rho ^\prime  /l_0}}{\rho ^\prime } d^2\mathbf{r}^\prime , \nn\\
\end{align}
where $\rho ^\prime = \sqrt{x^{\prime \, 2 } + y ^{\prime \, 2 }}$. This expression suggests the definition of the generic integrals (for $j = 0,\,1$)
\begin{align}
    \mathcal{I} _{j} (\rho) = \frac{1}{2 \pi l_0} \int d^2\mathbf{r}^\prime \frac{e^{- \rho ^{\prime} /l_0}}{ \rho ^{\prime} \sqrt{(x-x^{\prime})^{2} + (y-y^{\prime})^{2} + (2j z_{0})^{2} }} ,\label{Integral} 
\end{align}
such that the scalar potential contains both $\mathcal{I} _{0}$ and $\mathcal{I} _{1}$. To evaluate these integrals, we use the expansion: 
\begin{align}
&\frac{1}{\sqrt{(x-x^{\prime})^{2} + (y-y^{\prime})^{2} + (2j z_{0})^{2} }}\nn\\
 &= \sum_{m=-\infty}^{+\infty}\int_0^\infty dk~e^{\ii  m(\varphi-\varphi ^{\prime})}J_m(k\rho)J_m(k\rho ^{\prime})e^{-k (2j z_{0} )},\nn\\
\end{align}
so that
\begin{align}
    \mathcal{I} _{j}(\rho) &= \frac{1}{2 \pi l_0 } \sum_{m=-\infty}^{+\infty}\int_0^\infty dk \, J_m(k\rho) \, e^{-k (2j z_{0} )}\nn\\
    &\times \int _{0} ^{2 \pi } d \varphi ^{\prime} \, e^{\ii  m(\varphi-\varphi ^{\prime})} \int _{0} ^{\infty} d \rho^{\prime} \, J_m(k\rho ^{\prime}) \,  e^{-  \rho ^{\prime}/l_0}  . \nn\\
\end{align}
The angular integration yields $2 \pi \delta _{m0}$. Therefore
\begin{align}
    \mathcal{I} _{j}(\rho) = \frac{1}{l_0} \int_0^\infty dk \, J_{0} (k\rho)  e^{-k (2j z_{0} )} \! \int _{0} ^{\infty} d \rho^{\prime} \, J_{0}(k\rho ^{\prime})   e^{-  \rho ^{\prime}/l_0}  . 
\end{align}
Now, using the integral formula
\begin{align}
\int_{0}^{\infty} dx~e^{-\alpha x}J_{\nu}(\beta x)=\frac{\beta ^{- \nu} \left[ \sqrt{\alpha^2+\beta^2} - \alpha \right] ^{\nu} }{\sqrt{\alpha^2+\beta^2}} , \label{Int_Formula}
\end{align}
for $\mbox{Re} ( \alpha \pm i \beta ) > 0$ and $\mbox{Re} \, \nu > - 1$, we obtain
\begin{align}
    \mathcal{I} _{j}(\rho) = \int_0^\infty \!\! dk \, \frac{J_{0} (k\rho)}{\sqrt{1 + (k l _{0}) ^{2}}} \, e^{-k (2j z_{0} )} .
\end{align}
The case $j = 0$ has a closed form expression:
\begin{align}
    \mathcal{I} _{0}(\rho) = \int_0^\infty \!\! dk \, \frac{J_{0} (k\rho)}{\sqrt{1 + (k l _{0}) ^{2}}} = \frac{1}{l_0} \, I _{0} (\rho / 2l_0 ) \, K _{0} ( \rho / 2l_0 ) ,
\end{align}
where $I _{n}(z)$ and $K _{n}(z)$ are the modified Bessel functions of the first and second kind, respectively. The case $j=1$ can not be expressed in terms of simple functions, but it can easily be evaluated numerically. All in all, the above results fully provide the scalar potential of Eq.~\eqref{Eq:Scalarpotentialfinal}. For later use we further define the functions (for $j = 0,\,1$, respectively)
\begin{align}
    \Lambda ^{(\nu )} _{j} (\rho ) =  \int_0^\infty \!\! dk \, \frac{J_{\nu} (k\rho)}{\sqrt{1 + (k l _{0}) ^{2}}} \, e^{-k (2j z_{0} )} , \label{F_Lambda}
\end{align}
such that $\Lambda ^{(0 )} _{j} (\rho ) = \mathcal{I} _{j}(\rho)$.

We now turn to the evaluation of the vector potential, which is given by Eq. (\ref{VectPotential}) with the charge density of Eq. (\ref{SourceYukawa}), i.e. ${\bf{A}} ({\bf{r}}) = \int {\bf{G}} ({\bf{r}} , {\bf{r}}^{\prime}) \, \rho _\text{Y}({\bf{r}}^{\prime}) \, d^{3} {\bf{r}}^{\prime}$. To this end, we use the following integral identity \cite{PhysRevD.92.125015}:
\begin{align}
    & \frac{{\bf{R}}}{ R ^{2}} \left[ 1 - \frac{z+z^{\prime}}{ \sqrt{ R ^{2} + (z+z')^{2}} } \right] \notag \\[5pt] & \hspace{1.5cm} = \frac{1}{2 \pi i} \int  \frac{{\bf{k}}}{k^{2}} e ^{-k(z+z')} e ^{i {\bf{k}} \cdot  {\bf{R}} } d^{2} {\bf{k}} \notag \\[5pt]  & \hspace{1.5cm} = - \frac{1}{2 \pi } \nabla _{\perp} \int  \frac{1}{k^{2}} e ^{-k(z+z')} e ^{i {\bf{k}} \cdot  {\bf{R}} } \, d^{2} {\bf{k}} ,
\end{align}
where $\nabla _{\perp} = \hat{{\bf{e}}} _{x} \partial _{x} + \hat{{\bf{e}}} _{y} \partial _{y} $ is the gradient in the transverse coordinates and $ {\bf{R}} \equiv {\boldsymbol{\rho}} - {\boldsymbol{\rho}} ^{\prime} = (x-x^{\prime}) \hat{{\bf{e}}}_{x} + (y-y^{\prime}) \hat{{\bf{e}}}_{y} $. Using this result, the vector Green's function (\ref{Green_Vector}) can be expressed as
\begin{align}
    {\bf{G}} ({\bf{r}} , {\bf{r}}^{\prime}) = - \frac{\mu _{1} g}{8 \pi ^{2}} \, \hat{{\bf{e}}}_{z} \times \nabla _{\perp} \int  \frac{1}{k^{2}} e ^{-k(z+z')} e ^{i {\bf{k}} \cdot  {\bf{R}} } \, d^{2} {\bf{k}} ,
\end{align}
such that the vector potential becomes
\begin{widetext}
\begin{align}
    {\bf{A}} ({\bf{r}}) &= - \frac{\mu _{1} g}{8 \pi ^{2}} \, \hat{{\bf{e}}}_{z} \times \nabla _{\perp}  \, \int  \frac{d^{2} {\bf{k}}}{k ^{2}} \, e ^{i {\bf{k} } \cdot {\boldsymbol{\rho}}} \, e ^{-kz} \, \int d^{3} {\bf{r}}^{\prime} \, \rho _\text{Y} ({\bf{r}}^{\prime}) \, e ^{- i {\bf{k}} \cdot  {\boldsymbol{\rho}} ^{\prime}} \, e ^{-k z'} .
\end{align}
Substituting the charge density distribution Eq.~ (\ref{SourceYukawa}), and performing the integrals involving the Dirac deltas we obtain
\begin{align}
    {\bf{A}} ({\bf{r}}) &= - \frac{\mu _{1} Q g}{8 \pi ^{2}} \, \hat{{\bf{e}}}_{z} \times \nabla _{\perp}  \, \int  \frac{d^{2} {\bf{k}}}{k ^{2}} \, e ^{i {\bf{k} } \cdot {\boldsymbol{\rho}}} \, e ^{-k(z+z_{0})} \, \left[ 1 -\frac{1}{l_0} \int _{0} ^{\infty} d  \rho ^{\prime} \, J_{0} (k \rho ^{\prime}) \, e^{- \rho ^{\prime}/l_0 }  \right] .
\end{align}
The integral in $\rho^\prime$ can be easily evaluated by using the formula (\ref{Int_Formula}). 
To evaluate this integral we introduce polar coordinates such that ${\bf{k} } \cdot {\boldsymbol{\rho}} = k \rho \cos \varphi $. Performing the angular integration we obtain
\begin{align}
    {\bf{A}} ({\bf{r}}) &= - \frac{\mu _{1} Q g}{4 \pi } \, \hat{{\bf{e}}}_{z} \times \nabla _{\perp}  \, \int  \frac{dk}{k} \, J _{0} ( k \rho) \, e ^{-k(z+z_{0})} \, \left[ 1 - \frac{1}{\sqrt{1 + (k l_0 ) ^{2} }}  \right] .
\end{align}
Finally, taking the gradient, $- \nabla _{\perp}  J_{0}(k \rho) = k \, \hat{{\bf{e}}}_{\rho}  J_{1}(k \rho) $, and using that $\hat{{\bf{e}}}_{z} \times  \hat{{\bf{e}}}_{\rho} = \hat{{\bf{e}}}_{\phi}$, one finds
\begin{align}
    {\bf{A}} ({\bf{r}}) &= \frac{\mu _{1} Q g}{4 \pi } \, \hat{{\bf{e}}}_{\phi} \, \int _{0} ^{\infty}  dk \, J _{1} ( k \rho) \, e ^{-2k z_{0}} \, \left[ 1 - \frac{1}{\sqrt{1 + (k l_0 ) ^{2} }}  \right] .
\end{align}
The first integral can be evaluated by using the integral formula (\ref{Int_Formula}) and yields the Schwinger vector potential for a magnetic monopole at the image point $- z_{0}$. It corresponds to the first term in Eq. (\ref{Eq:VectorPotential}). The second term can not be expressed in terms of simple functions, however it corresponds to the function $\Lambda ^{(1)} _{1} (\rho )$, defined by Eq. (\ref{F_Lambda}). On the whole we get the expression (\ref{Eq:VectorPotential}) for the vector potential.

\end{widetext}

\section{The scattered states}\label{sec:GreensFunctionCalculation}
\subsection{The Lippman-Schwinger equation and free Green's function}

In this section, we shall closely follow the formalism presented in Ref.~\cite{nano12203711}.
In the elastic scattering theory, we look for spinor solutions $\left|\Psi_{\mu,\mathbf{k}}\right\rangle$ of the total Hamiltonian of Eq.~\eqref{eq:complete_H} with the same energy as in Eq.~\eqref{eq:unpert_spectrum}. That solution is given by the well known Lippmann-Schwinger equation
\begin{equation}
	\left| \Psi_{\mathbf{k},\mu}\right\rangle = \left|\Phi_{\mathbf{k},\mu}\right\rangle  + \hat{G}^{\xi}_{R,0}(E)\hat{H}_{1}^{\xi} \left| \Psi_{\mathbf{k},\mu}\right\rangle, \label{eq:Lippmann_Schwinger}
\end{equation}
where the Green's operator $\hat{G}^{\xi}_{R,0}(E)$ or {\it resolvent} is given by
\begin{equation}
	\hat{G}^{\xi}_{R,0}(E)=\frac{1}{E-\hat{H}^{\xi}_0+ \ii\eta^{+}}, \label{eq:resolvent}
\end{equation}
where the positive sign for the regulator $\ii\eta^+$ defines the retarded Green's function, which in turn produces outgoing spherical waves from the scattering center. Moreover, the advanced Green's function is defined as 
\begin{equation}
	\hat{G}^{\xi}_{A,0}(E) =\left[ \hat{G}^{\xi}_{R,0}(E)\right] ^{\dagger},\label{eq:relation_GR_GA}
\end{equation}
and using the explicit form of the resolvent of Eq.~\eqref{eq:resolvent} we have a relation between the retarded and the advanced GFs
\begin{equation}
	\hat{G}^{\xi}_{R,0}(E)-\hat{G}^{\xi}_{A,0}(E)=-2\pi \ii \delta\left(E-\hat{H}^{\xi}_0\right).\label{eq:GR-GA}
\end{equation}

The resolvent in Eq.~\eqref{eq:resolvent} is the solution to the equation
\begin{equation}
	\left( E + \ii\eta^{+}-\hat{H}^{\xi}_0\right) \hat{G}^{\xi}_{R,0}(E)=\hat{I}. \label{eq:eq_resolvent_0}
\end{equation}

At this point we introduce the $\hat{T}$ matrix as usual
\begin{equation}
	\hat{T}^{\xi}(E)\left| \Phi_{\mathbf{k},\mu}\right\rangle=\hat{H}_{1}^{\xi} \left| \Psi_{\mathbf{k},\mu}\right\rangle,
	\label{eq:T_matrix_def}
\end{equation}
so that the Lippmann-Schwinger equation becomes 
\begin{equation}
	\left| \Psi_{\mathbf{k},\mu}\right\rangle = \left(\hat{I}  + \hat{G}^{\xi}_{R,0}(E)\hat{T}^{\xi}(E)\right) \left| \Phi_{\mathbf{k},\mu}\right\rangle. \label{eq:Lippmann_T_matrix}
\end{equation}

Inserting Eq.~\eqref{eq:Lippmann_T_matrix} into Eq.~\eqref{eq:T_matrix_def}, and solving for the $\hat{T}$-matrix operator, we find the formal expression
\begin{equation}
	\hat{T}^{\xi}(E)= \hat{H}_{1}^{\xi}\left(\hat{I}- \hat{G}^{\xi}_{R,0}(E)\hat{H}_{1}^{\xi}\right) ^{-1}.
	\label{eq:T1}
\end{equation}

We are also interested in the total retarded Green's function, which is the solution to the equation
\begin{equation}
	\left( E + \ii\eta^{+}-\hat{H}^{\xi}\right) \hat{G}^{\xi}_{R}(E)=\hat{I},
\end{equation}
where $\hat{H}^{\xi}=\hat{H}^{\xi}_0+\hat{H}^{\xi}_1$ is the full Hamiltonian. The last equation can be transformed into a self-consistent relation
\begin{equation}
	\left( E + \ii\eta^{+}-\hat{H}^{\xi}_0\right) \hat{G}^{\xi}_{R}(E)=\hat{I}+\hat{H}^{\xi}_1\hat{G}^{\xi}_{R}(E),
\end{equation}
and making use of  Eq.~\eqref{eq:eq_resolvent_0} we have
\begin{equation}
	\hat{G}^{\xi}_{R}(E)=\hat{G}^{\xi}_{R,0}(E)+\hat{G}^{\xi}_{R,0}(E)\hat{H}^{\xi}_1\hat{G}^{\xi}_{R}(E).
\end{equation}

The formal solution of this equation is
\begin{equation}
	\hat{G}^{\xi}_{R}(E) = \left[\hat{I} -  \hat{G}^{\xi}_{R,0}(E)\hat{H}^{\xi}_1\right]^{-1}\hat{G}^{\xi}_{R,0}(E),
\end{equation}
that combined with Eq.~\eqref{eq:T1} leads to the identity
\begin{eqnarray}
	\hat{H}_1^{\xi} \hat{G}^{\xi}_{R}(E) = \hat{T}^{\xi}(E) \hat{G}^{\xi}_{R,0}(E).
\end{eqnarray}

Then, the complete retarded GF is given by
\begin{equation}
	\hat{G}^{\xi}_{R}(E)=\hat{G}^{\xi}_{R,0}(E)+\hat{G}^{\xi}_{R,0}(E)\hat{T}^{\xi}(E)\hat{G}^{\xi}_{R,0}(E), \label{eq:Total_G_T_matrix}
\end{equation}
and hence one can show that the $T$-matrix itself satisfies a self-consistent equation of the form
	\begin{equation}
		\hat{T}^{\xi}(E)=  \hat{H}_1^{\xi}+\hat{H}_1^{\xi}\hat{G}^{\xi}_{R,0}(E) \hat{T}^{\xi}(E).
	\end{equation}
	
	From the latter it follows that
	\begin{equation}
		\hat{T}^{\xi}(E)-\left[\hat{T}^{\xi}(E)\right]^{\dagger}=\left[\hat{T}^{\xi}(E)\right]^{\dagger}\left(\hat{G}^{\xi}_{R}(E)-\hat{G}^{\xi}_{A}(E)\right)\hat{T}^{\xi}(E),
	\end{equation}
	and using the Eq.~\eqref{eq:GR-GA} we get an useful expression:
	\begin{equation}
		\hat{T}^{\xi}(E)-\left[\hat{T}^{\xi}(E)\right]^{\dagger}=-2\pi \ii\left[\hat{T}^{\xi}(E)\right]^{\dagger}\delta\left(E-\hat{H}^{\xi}_0\right)\hat{T}^{\xi}(E).\label{eq:T-Tdagger}
\end{equation}

\begin{figure}
	\centering
	\includegraphics[width=1\columnwidth]{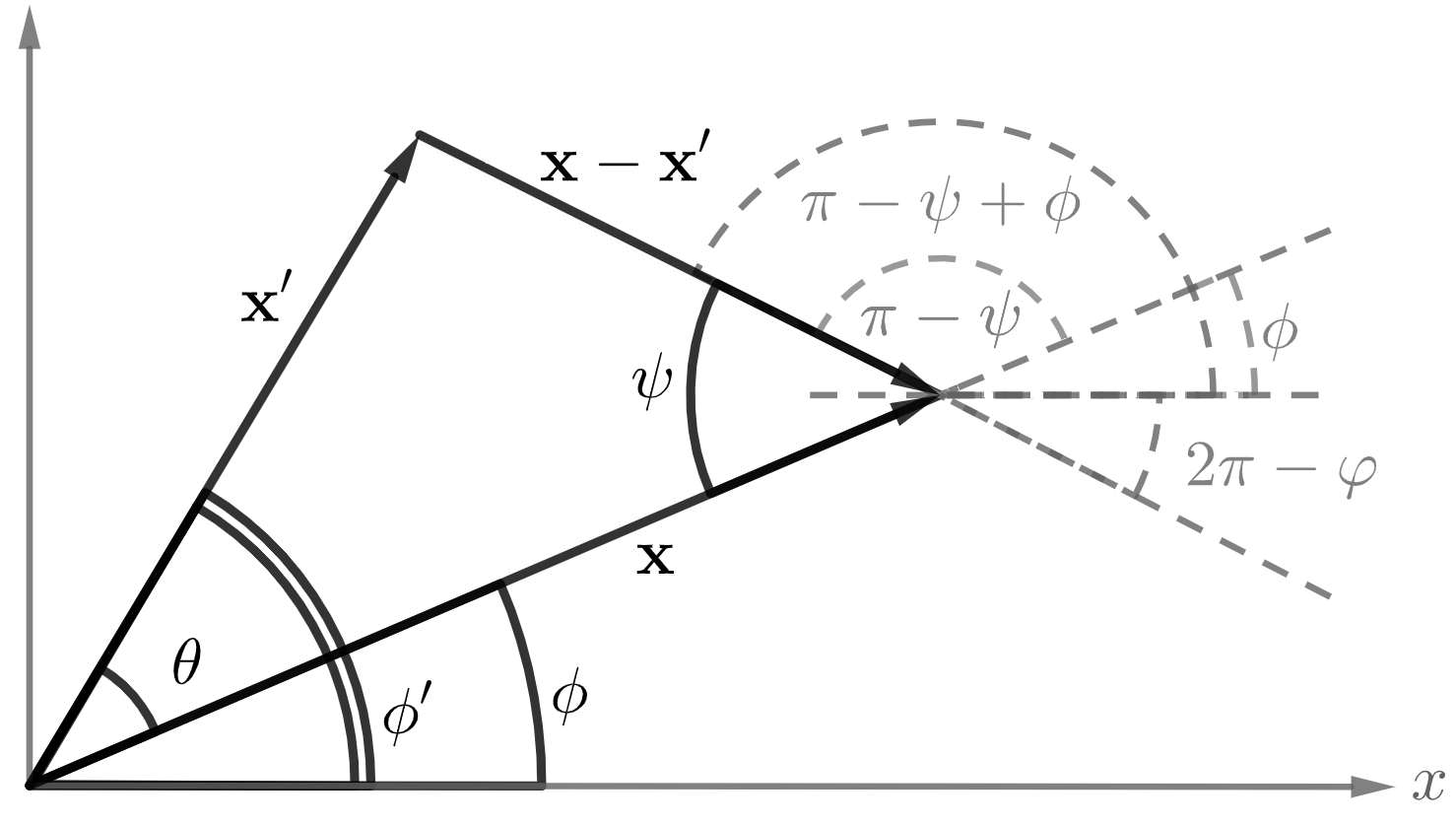}
	\caption{Triangle used for the application of the addition theorem for the Bessel functions.}
	\label{fig:coord_add_thrm}
\end{figure}

The form for the retarded Green's function matrix in the coordinates representation is \cite{nano12203711}
\begin{widetext}
\begin{equation}
	\mathbf{G}^{\xi }_{R,0}\left( \mathbf{x},\mathbf{x'};k\right)= -\frac{\lambda \xi \ii k}{4 \hbar v_\text{F}}\begin{bmatrix}
		H^{(1)}_0 \left(k|\mathbf{x}-\mathbf{x'}| \right) &  \ii\lambda e^{-\ii\varphi} H^{(1)}_1 \left(k|\mathbf{x}-\mathbf{x'}|\right) \\
		\ii\lambda e^{\ii\varphi} H^{(1)}_1 \left(k|\mathbf{x}-\mathbf{x'}|\right) &  H^{(1)}_0 \left(k|\mathbf{x}-\mathbf{x'}| \right)
	\end{bmatrix}, \label{eq:G0_final}
	\end{equation}
\end{widetext}
where $\varphi$ is the angle of the vector $\mathbf{x}-\mathbf{x'}$ w.r.t. the $x$ axis.  We want to expand this matrix Green´s function in polar coordinates. In order to do it, we use the addition theorem for Bessel functions~\cite{Abramowitz}
\begin{equation}
	e^{\pm \ii \nu \psi}Z_{\nu}(\lb R) = \sum_{m=-\infty}^{\infty}Z_{\nu+m}(\lb r_2)J_m(\lb r_1)e^{\pm \ii m\theta},\label{eq:addition_theorem} 
\end{equation}
where $Z_{\nu}$ is any of the Bessel functions $J_{\nu}$, $Y_{\nu}$, $H_{\nu}^{(1)}$, or $H_{\nu}^{(2)}$. The quantities $r_1$, $r_2$, $R$ are given by $\mathbf{x}$, $\mathbf{x}^\prime$ and $|\mathbf{x}-\mathbf{x}^\prime|$, respectively. Moreover, the angles $\theta$ and $\psi$ are defined in the triangle of Fig.~\ref{fig:coord_add_thrm}, from which
\bea
\varphi=2\pi+\phi-\psi.
\eea

Let us write the Green's function as:
\begin{equation}
	\mathbf{G}^{\xi }_{R,0}\left( \mathbf{x},\mathbf{x'};k\right)= -\frac{\lambda \xi \ii k}{4 \hbar v_\text{F}}\begin{bmatrix}
		G_{11}& G_{12}\\
		G_{21}& G_{22}
	\end{bmatrix},
\end{equation} 
so that by using
\begin{align}
H^{(1)}_0\left(k|\mathbf{x}-\mathbf{x'}| \right)
	&= \sum_{m=-\infty}^{\infty}H_{m}^{(1)}(k r)J_m(k r')e^{\ii  m(\f'-\f)}\notag \\
	&= \sum_{m=-\infty}^{\infty}H_{-m}^{(1)}(k r)J_{-m}(k r')e^{- i m(\f'-\f)},
	\label{Eq:H01}
\end{align}
the components $G_{11}$, and $G_{22}$ are obtained by shifting $m$ Eq.~\eqref{Eq:H01} as $m\to-m$ and $m\to m+1$, respectively. The latter, together with the Bessel function's properties yields
\bea
G_{11}=\sum_{m=-\infty}^{\infty}H_{m}^{(1)}(k r)J_{m}(k r')e^{ i m(\f-\f')},
\eea
and
\bea
G_{12}=\sum_{m=-\infty}^{\infty}H_{m+1}^{(1)}(k r)J_{m+1}(k r')e^{ i (m+1)(\f-\f')}.\nn\\
\eea

On the other hand:
\begin{align}
	G_{21}&=\ii\lb e^{\ii \varphi}H^{(1)}_1 \left(k|\mathbf{x}-\mathbf{x'}| \right) \notag\\
	&=\ii\lb e^{\ii (2\p +\f -\psi)}H^{(1)}_1 \left(k|\mathbf{x}-\mathbf{x'}| \right) \notag\\
	&=\ii\lb e^{\ii \f }e^{-\ii \psi}H^{(1)}_1 \left(k|\mathbf{x}-\mathbf{x'}| \right)  \notag \\
	&=\ii\lb e^{\ii \f } \sum_{m=-\infty}^{\infty}H_{m+1}^{(1)}(k r)J_m(k r')e^{-\ii  m(\f'-\f)}\notag \\
	&= \ii\lb \sum_{m=-\infty}^{\infty}H_{m+1}^{(1)}(k r)J_m(k r')e^{\ii \f } e^{\ii  m(\f-\f')}.
\end{align}
and
\begin{align}
	G_{12}&=\ii\lb e^{-\ii \varphi}H^{(1)}_1 \left(k|\mathbf{x}-\mathbf{x'}| \right) \notag\\
	&=\ii\lb e^{-\ii (2\p +\f -\psi)}H^{(1)}_1 \left(k|\mathbf{x}-\mathbf{x'}| \right)\notag\\
	&=\ii\lb e^{-\ii \f } \sum_{m=-\infty}^{\infty}H_{m+1}^{(1)}(k r)J_m(k r')e^{\ii  m(\f'-\f)}\notag \\
	&= \ii\lb \sum_{m=-\infty}^{\infty}H_{-m}^{(1)}(k r)J_{-m-1}(k r')e^{-\ii \f } e^{-\ii  (-m-1)(\f-\f')}\notag \\
	&= -\ii\lb \sum_{m=-\infty}^{\infty}H_{m}^{(1)}(k r)J_{m+1}(k r')e^{-\ii \f '} e^{\ii m(\f-\f')},
\end{align}
where in the last step we shifted $m\to-m-1$.

Then, if we consider that $r<r'$
\begin{widetext}
\begin{align}
	\mathbf{G}^{\xi }_{R,0}\left( \mathbf{x},\mathbf{x'};k\right)=-\frac{\lb\xi \ii k}{4\hbar v_\text{F}}\sum_{m\in\mbb{Z}}\begin{bmatrix}
		J_m(kr_{<})H_m^{(1)}(kr_{>})e^{\ii m(\f-\f')} & -\ii\lb J_{m+1}(kr_{<})H_m^{(1)}(kr_{>}) e^{-\ii \f'}e^{\ii m(\f-\f')} \\
		\ii \lb J_{m}(kr_{<})H_{m+1}^{(1)}(kr_{>}) e^{\ii \f}e^{\ii m(\f-\f')} & J_{m+1}(kr_{<})H_{m+1}^{(1)}(kr_{>}) e^{\ii (m+1)(\f-\f')} 
	\end{bmatrix},\label{eq:Green_Fn_final}
\end{align}
where $r_{>}$ ($r_{<}$) is the greater (lower) between $r$ and $r'$. 
\end{widetext}

\subsection{The radial integral equation}

Now, representing the Lippmann-Schwinger  Eq.~\eqref{eq:Lippmann_Schwinger} in the coordinate basis we have
\bea
	\left\langle \mathbf{x}\middle|\Psi_{\mathbf{k},\lambda}\right\rangle &=& \left\langle \mathbf{x}\middle|\Phi_{\mathbf{k},\lambda}\right\rangle\nn\\
	&+&\int _{\mbb{R}^2}d^2x'  \mathbf{G}^{\xi }_{R,0}\left( \mathbf{x},\mathbf{x'};k\right) \mathbf{H}^{\xi}_1(\mathbf{x'})	\left\langle \mathbf{x'} \middle|\Psi_{\mathbf{k},\lambda}\right\rangle,\nn\\
\eea
where $\left\langle \mathbf{x}\middle|\Psi_{\mathbf{k},\lambda}\right\rangle$ is given in Eq.~\eqref{eq:full_spinor_angular}, $\left\langle \mathbf{x}\middle|\Phi_{\mathbf{k},\lambda}\right\rangle$ is the free spinor of Eq.~\eqref{eq:incident_spinor_angular}, $\mathbf{G}^{\xi }_{R,0}\left( \mathbf{x},\mathbf{x'};k\right)$ is the matrix Green's function given in Eq.~\eqref{eq:Green_Fn_final}, and the matrix form of the operator representing the interaction with the external fields $\mathbf{H}_1^{\xi}$ is
\begin{equation}
	\mathbf{H}^{\xi}_1(\mathbf{x})=-\xi q v_\text{F} \boldsymbol{\sigma}\cdot\boldsymbol{\hat{e}}_{\phi} A(\mathbf{r})+q\Phi(\mathbf{r})\sigma_0,
\end{equation}
where the scalar potential $\Phi(\mathbf{r})$ is given in Eq.~\eqref{Eq:Scalarpotentialfinal}, the magnitude of the vector potential $\mathbf{A}(\mathbf{r}) = \boldsymbol{\hat{e}}_{\phi} A(\mathbf{r})$ is given in Eq.~\eqref{Eq:VectorPotential}, $\sigma_0$ is the $2\times2$ unit matrix and 
\begin{equation}
\boldsymbol{\sigma}\cdot \boldsymbol{\hat{e}}_{\phi}=\begin{bmatrix}
	0 & -\ii e^{-\ii\phi} \\
	\ii e^{\ii\phi} & 0
\end{bmatrix}.
\end{equation}
Then, we have
\begin{equation}
	\mathbf{H}^{\xi}_1(\mathbf{x})=\begin{bmatrix}
		q\Phi(r) & \ii q^{-\ii\phi} \xi  qv_\text{F} A(r)\\
		-\ii q^{\ii\phi} \xi  qv_\text{F} A(r) & q\Phi(r)
	\end{bmatrix}.
\end{equation}

After we make all the replacements,  we can extract the radial part to obtain the radial integral equation. For $r<a$ the final result can be written as
\begin{widetext}
{\small
\begin{align}
\begin{bmatrix}
	f_m(r) \\ g_m(r)
\end{bmatrix}&=\begin{bmatrix}
J_m(kr) \\  J_{m+1}(kr)
\end{bmatrix} -\frac{\lb\xi \ii \p  k}{2\hbar v_\text{F}}\int_{0}^{r} dr' r' \begin{bmatrix}
J_m(kr')H_m^{(1)}(kr)& J_{m+1}(kr')H_m^{(1)}(kr) \\
 J_{m}(kr')H_{m+1}^{(1)}(kr) & J_{m+1}(kr')H_{m+1}^{(1)}(kr) 
\end{bmatrix} \begin{bmatrix}
e\Phi(r')& -\lb\xi qv_\text{F} A(r') \\
-\lb\xi qv_\text{F} A(r') & e\Phi(r')
\end{bmatrix}\begin{bmatrix}
f_m(r') \\ g_m(r')
\end{bmatrix}\notag \\
&\quad -\frac{\lb\xi \ii \p  k}{2\hbar v_\text{F}}\int_{r}^{a} dr' r' \begin{bmatrix}
J_m(kr)H_m^{(1)}(kr')&  J_{m+1}(kr)H_m^{(1)}(kr') \\
 J_{m}(kr)H_{m+1}^{(1)}(kr') & J_{m+1}(kr)H_{m+1}^{(1)}(kr') 
\end{bmatrix} \begin{bmatrix}
e\Phi(r')&- \lb \xi qv_\text{F} A(r') \\
- \lb \xi qv_\text{F} A(r') & e\Phi(r')
\end{bmatrix}\begin{bmatrix}
f_m(r') \\ g_m(r')
\end{bmatrix}. \label{eq:integral_eqn_final}
\end{align}
}
and for $r>a$  we have
{\small
\begin{align}
&\begin{bmatrix}
	f_m(r) \\ g_m(r)
\end{bmatrix}=\begin{bmatrix}
J_m(kr) \\  J_{m+1}(kr)
\end{bmatrix}-\frac{\lb\xi \ii \p  k}{2\hbar v_\text{F}}\int_{0}^{a} dr' r' \begin{bmatrix}
J_m(kr')H_m^{(1)}(kr)&  J_{m+1}(kr')H_m^{(1)}(kr) \\
 J_{m}(kr')H_{m+1}^{(1)}(kr) & J_{m+1}(kr')H_{m+1}^{(1)}(kr) 
\end{bmatrix} \begin{bmatrix}
e\Phi(r')& - \lb \xi ev_\text{F} A(r') \\
- \lb \xi ev_\text{F} A(r') & e\Phi(r')
\end{bmatrix}\begin{bmatrix}
f_m(r') \\ g_m(r')
\end{bmatrix} .\label{eq:integral_eqn_final_2}
\end{align}
}
\end{widetext}

Finally, in order to obtain asymptotic eigenstates, we expand the Bessel's functions when $x\rightarrow\infty $ as follows:
\begin{align}
	J_{\nu}(x)&\rightarrow\sqrt{\frac{1}{2\p x}} \left[ e^{ \ii\left( x-\frac{\nu \pi}{2}-\frac{\p}{4} \right) }+e^{- \ii\left( x-\frac{\nu \pi}{2}-\frac{\p}{4} \right) }\right] ,\nn\\
	H^{(1)}_{\nu}(x)&\rightarrow\sqrt{\frac{2}{\p x}}e^{ \ii\left( x-\frac{\nu \pi}{2}-\frac{\p}{4} \right) }.
\end{align}

Then
\begin{equation}
\begin{bmatrix}
	J_m(kr) \\J_{m+1}(kr)
\end{bmatrix}\rightarrow \sqrt{\frac{2}{\p kr}}\sum_{s=\pm 1} \begin{bmatrix}
1 \\ \mp \ii s
\end{bmatrix}e^{\pm \ii s\left(kr-\frac{m \pi}{2}-\frac{\p}{4}\right) },
\end{equation}
and
\bea
&&\begin{bmatrix}
	J_m(kr')H_m^{(1)}(kr)&  J_{m+1}(kr')H_m^{(1)}(kr) \\
 J_{m}(kr')H_{m+1}^{(1)}(kr) & J_{m+1}(kr')H_{m+1}^{(1)}(kr) 
\end{bmatrix} \notag\\
&\rightarrow& \sqrt{\frac{2}{\p k }}\begin{bmatrix}
	1 \\- \ii
\end{bmatrix} \begin{bmatrix}
J_m(kr') &  J_{m+1} (kr')
\end{bmatrix} \frac{e^{\ii\left(kr-\frac{m \pi}{2}-\frac{\p}{4}\right) }}{\sqrt{r}} .\nn\\
\eea
so that after replacing in Eq.~\eqref{eq:integral_eqn_final_2}, we obtain the result shown in Eq.~\eqref{eq:phase_shift_final}.

\section{The transport coefficients}\label{sec:The transport coefficients}
In this section, we shall present the mathematical details involved in the calculation of the electronic transport coefficients, in particular the electrical conductivity, leading to the results presented in the main body of the article.
\subsection{The Onsager's coefficients}\label{sec:The Onsager's coefficients}

In order to compute the microscopic transport coefficients, let us start by defining the quantum operators representing the particle, energy and heat currents, respectively:
 \begin{equation}
 \hat{\mathbf{j}}=\sum_{\mathbf{p}\sigma}\mathbf{v}_{\mathbf{p}}\hat{n}_{\mathbf{p}\sigma},\label{eq:current_operator}
 \end{equation}
  \begin{equation}
 \hat{\mathbf{j}}_{E}=\sum_{\mathbf{p}\sigma}\mathbf{v}_{\mathbf{p}}\mathcal{E}_{\mathbf{p}\sigma}\hat{n}_{\mathbf{p}\sigma},\label{eq:current_energy_operator}
 \end{equation}
\begin{equation}
 \hat{\mathbf{j}}_{Q}=\sum_{\mathbf{p}\sigma}\mathbf{v}_{\mathbf{p}}\left(\mathcal{E}_{\mathbf{p}\sigma}-\mu\right)\hat{n}_{\mathbf{p}\sigma},\label{eq:current_heat_operator}
 \end{equation}
 where for a particle with momentum $\mathbf{p}$ and spin $\sigma$, we identify $\mathcal{E}_{\mathbf{p}\sigma}$ as the particle's energy, $\mathbf{v}_{\mathbf{p}} = -\nabla_{\mathbf{p}}\mathcal{E}_{\mathbf{p}\sigma}$ as the group velocity at momentum $\mathbf{p}$, $\hat{n}_{\mathbf{p}\sigma}$ as the particle number density operator, and $\mu$ as the chemical potential of the system. The corresponding observed macroscopic currents are given by the ensemble average of the above operators, i.e.,
  \begin{subequations}
 \begin{align}
     \mathbf{J}&=\left\langle  \hat{\mathbf{j}} \right\rangle,\\
     \mathbf{J}_{E}&=\left\langle  \hat{\mathbf{j}}_{E} \right\rangle,\\
     \mathbf{J}_{Q}&=\left\langle  \hat{\mathbf{j}}_{Q} 
     \right\rangle.
     \label{eq:current_averages}
 \end{align}
 \end{subequations}

Now,  let us introduce the Onsager's coefficients $L_{\alpha\beta}^{(ij)}$ through the relations linking the currents with the gradients in temperature and electro-chemical potential \cite{Munoz_2012}
\begin{subequations}
    \begin{align}
     J_{\alpha}&=-\frac{1}{T}L_{\alpha\beta}^{(11)}\nabla_{\beta}\left(\mu+eV\right)+L_{\alpha\beta}^{(12)}\nabla_{\beta}\left(\frac{1}{T}\right),\\
     J_{Q\alpha}&=-\frac{1}{T}L_{\alpha\beta}^{(21)}\nabla_{\beta}\left(\mu+eV\right)+L_{\alpha\beta}^{(22)}\nabla_{\beta}\left(\frac{1}{T}\right),
 \label{eq:def_Onsager_coeff}
 \end{align}
\end{subequations}
which  in tensor notation take the form
 \begin{subequations}
     \begin{align}
     \mathbf{J}&=-\frac{1}{T}\overleftrightarrow{\mathbf{L}}^{(11)}\cdot\nabla\left(\mu+eV\right)+\overleftrightarrow{\mathbf{L}}^{(12)}\cdot\nabla\left(\frac{1}{T}\right),\label{eq:def_Onsager_coeff_Tensor_a}\\
     \mathbf{J}_{Q}&=-\frac{1}{T}\overleftrightarrow{\mathbf{L}}^{(21)}\cdot\nabla\left(\mu+eV\right)+\overleftrightarrow{\mathbf{L}}^{(22)}\cdot\nabla\left(\frac{1}{T}\right). \label{eq:def_Onsager_coeff_Tensor_b}
 \end{align}
 \end{subequations}

 The above tensors are directly related to the transport coefficients, as can be shown by exploring different limits\cite{Munoz_2012}. First, note that for $\nabla T=0$ and $\nabla\mu=0$, the current is purely electric. Therefore:
  \begin{equation}
     e\mathbf{J}=-\frac{1}{T}\overleftrightarrow{\mathbf{L}}^{(11)}\cdot\nabla\left(e^2 V\right)=\overleftrightarrow{\boldsymbol{\sigma}}\cdot\left(-\nabla V\right),
 \end{equation}
and the electrical conductivity tensor is given by
 \begin{equation}
\overleftrightarrow{\boldsymbol{\sigma}}=\frac{e^2}{T}\overleftrightarrow{\mathbf{L}}^{(11)}.\label{eq:conductivity_tensor}
 \end{equation}

 The same procedure is applied to find the thermal conductivity, which is given by the condition $\mathbf{J}=0$. Thus, from Eq.~\eqref{eq:def_Onsager_coeff_Tensor_a} we have
 \begin{equation}
     \mathbf{0}=-\frac{1}{T}\overleftrightarrow{\mathbf{L}}^{(11)}\cdot\nabla\left(\mu+eV\right)+\overleftrightarrow{\mathbf{L}}^{(12)}\cdot\nabla\left(\frac{1}{T}\right),
 \end{equation}
 and solving we obtain
 \begin{equation}
     \frac{1}{T}\nabla\left(\mu+eV\right)=\left[\overleftrightarrow{\mathbf{L}}^{(11)}\right]^{-1}\cdot\overleftrightarrow{\mathbf{L}}^{(12)}\cdot \nabla \left(\frac{1}{T}\right).
 \end{equation}

By substituting in Eq.~\eqref{eq:def_Onsager_coeff_Tensor_b} we have
 \begin{align}
    \mathbf{J}_{Q}&=-\overleftrightarrow{\mathbf{L}}^{(21)}\cdot\left[\overleftrightarrow{\mathbf{L}}^{(11)}\right]^{-1}\cdot\overleftrightarrow{\mathbf{L}}^{(12)}\cdot\nabla\left(\frac{1}{T}\right)\nn\\
    &+\overleftrightarrow{\mathbf{L}}^{(22)}\cdot\nabla\left(\frac{1}{T}\right)\notag \\
    &=-\frac{1}{T^2}\left(\overleftrightarrow{\mathbf{L}}^{(22)}-\overleftrightarrow{\mathbf{L}}^{(21)}\cdot\left[\overleftrightarrow{\mathbf{L}}^{(11)}\right]^{-1}\cdot\overleftrightarrow{\mathbf{L}}^{(12)}\right)\cdot \nabla T .
 \end{align}

Then, the thermal conductivity tensor is
 \begin{equation}
     \overleftrightarrow{\boldsymbol{\kappa}}=\frac{1}{T^2}\left(\overleftrightarrow{\mathbf{L}}^{(22)}-\overleftrightarrow{\mathbf{L}}^{(21)}\cdot\left[\overleftrightarrow{\mathbf{L}}^{(11)}\right]^{-1}\cdot\overleftrightarrow{\mathbf{L}}^{(12)} \right),\label{eq:thermal_conductivity_tensor}
 \end{equation}
and the Seebeck coefficient (thermopower) is given by
 \begin{equation}
     S=\frac{1}{eT}\left[\overleftrightarrow{\mathbf{L}}^{(11)}\right]^{-1}\cdot\overleftrightarrow{\mathbf{L}}^{(12)}.\label{eq:Seebeck}
 \end{equation}

\subsection{The Linear Response Regime}\label{The Linear Response Regime}
To express the Onsager's coefficients in terms of dynamical variables, we apply the Kubo formalism within the linear response regime. For that sake, we express the entropy production rate as\cite{Munoz_2012}
 \begin{align}
     \frac{d Q}{d t}&=T\frac{\partial S}{\partial t}\nn\\
     &= -\mathbf{J}\cdot\nabla\left(\mu+eV\right)+T\,\mathbf{J}_{Q}\cdot\nabla\left(\frac{1}{T}\right)\notag\\
     &=T\sum_{i}\mathbf{J}_{i}\cdot \mathbf{X}_{i}\nn\\
     &\equiv \frac{\partial }{\partial t}F(t),\label{eq:dF_dt}
 \end{align}
 where $\mathbf{J}_{1} = \mathbf{J}$, $\mathbf{J}_{2}=\mathbf{J}_{Q}$, and 
 \begin{align}
     \begin{split}
         \mathbf{X}_{1}&=-\frac{1}{T}\nabla \left(\mu+eV\right),\\
         \mathbf{X}_{2}&=\nabla \left(\frac{1}{T}\right).
     \end{split}
 \end{align}

  We know that the Kubo's formula is given by\cite{Mahan}
 \begin{equation}
     \mathbf{J}_{i}=-\int_{0}^{\infty}dt\,e^{-s\,t}\int_{0}^{\beta}d\beta'\,\Tr \left[\hat{\rho}_{0}\frac{\partial }{\partial t}F(-t-\ii\,\hbar \beta')\hat{\mathbf{j}}_{i}(\mathbf{x})\right],\label{eq:Kubo_1}
 \end{equation}
 where $s$ is a positive quantity that guarantees the adiabatic switching-on of the perturbation, so that at the end of the calculation, we take the limit $s\rightarrow0$. When inserting Eq.~\eqref{eq:dF_dt} into Eq.~\eqref{eq:Kubo_1}, we have
{\small
    \begin{align}
             \mathbf{J}_{i}=-T\mathcal{L}\left\{\int_{0}^{\beta}d\beta'\,\Tr \left[\hat{\rho}_{0}\left(\sum_{k}\hat{\mathbf{j}}_{k}(-t-\ii\hbar \beta')\cdot\mathbf{X}_{k}\right)\hat{\mathbf{j}}_{i}(\mathbf{x})\right]\right\},\label{eq:Kubo_2}
    \end{align}
    }
 with $\mathcal{L}\{\cdot\}$ the Laplace's transform in the variable $s$:
 \bea
 \laplace{f(t)}=\int_0^\infty dt\,e^{-st}f(t).
 \eea
 
 Then, 
 \begin{subequations}
    \begin{align}
     L_{\alpha\beta}^{(11)}&=-T\laplace{\int_{0}^{\beta}d\beta'\,\Tr \left[\hat{\rho}_{0}\hat{j}_{\alpha}(-t-\ii\hbar \beta')\hat{j}_{\beta}\right]},\label{eq:L_11_Kubo}\\
     L_{\alpha\beta}^{(12)}&=L_{\alpha\beta}^{(21)}=-T\laplace{\int_{0}^{\beta}d\beta'\,\Tr \left[\hat{\rho}_{0}\hat{j}_{Q\alpha}(-t-\ii\hbar \beta')\hat{j}_{\beta}\right]},\label{eq:L_12_Kubo}\\
     L_{\alpha\beta}^{(22)}&=-T\laplace{\int_{0}^{\beta}d\beta'\,\Tr \left[\hat{\rho}_{0}\hat{j}_{Q\alpha}(-t-\ii\hbar \beta')\hat{j}_{Q\beta}\right]}\label{eq:L_22_Kubo}.
 \end{align} 
 \end{subequations}

To compute the traces, we define the current operator by
\begin{equation}
    \hat{\mathbf{j}}^{(\xi)}(\mathbf{x})=\xi v_\text{F} \left|\mathbf{x}\middle\rangle\boldsymbol{\sigma}\middle\langle\mathbf{x}\right|,\label{eq:electric_current_op}
\end{equation}
the heat current operator
\begin{equation}
    \hat{\mathbf{j}}^{(\xi)}_{Q}(\mathbf{x})=\xi v_\text{F} (\hat{H}^{\xi}-\mu)\left|\mathbf{x}\middle\rangle\boldsymbol{\sigma}\middle\langle\mathbf{x}\right|,\label{eq:heat_current_op}
\end{equation}
and the equilibrium density operator
\begin{equation}
    \hat{\rho}_0=\frac{\exp\left[-\beta \left(\hat{H}^{\xi}-\mu\right)\right]}{\mathcal{Z}(\beta,V,\mu)},\label{eq:density_operator}
\end{equation}
where $\mathcal{Z}(\beta,V,\mu)=\Tr \exp\left[-\beta \left(\hat{H}^{\xi}-\mu\right)\right]$ is the grand-canonical partition function. Here, $\hat{H}^{\xi}$ is the Dirac Hamiltonian with chirality $\xi$ and Fermi's velocity $v_\text{F}$. 

We start with Eq.~\eqref{eq:L_11_Kubo} and take the trace in the complete and orthonormal basis $\left\lbrace\left| \Psi_{\lambda,\mathbf{k}_{\parallel}}\right\rangle\right\rbrace$ of the total Hamiltonian, such that
\bea
\hat{H}^{\xi}\left| \Psi_{\lambda,\mathbf{k}_{\parallel}}\right\rangle =\mathcal{E}^{\lambda,\xi}_{\mathbf{k}_{\parallel}}\left| \Psi_{\lambda,\mathbf{k}_{\parallel}}\right\rangle,
\eea
from which
\begin{widetext}
    \begin{align}
L_{\alpha\beta}^{(11)}(\mathbf{x},\mathbf{x'})&=-T\laplace{\int_{0}^{\beta}d\beta'\,\Tr \left[\hat{\rho}_{0}\hat{j}_{\alpha}(\mathbf{x},-t-\ii\hbar \beta')\hat{j}_{\beta}(\mathbf{x'},0)\right]}\notag\\
&=-T\laplace{\int_{0}^{\beta}d\beta'\,\Tr \left[\hat{\rho}_{0}e^{\ii(-t-\ii\hbar \beta')\hat{H}^{\xi}/\hbar}\hat{j}_{\alpha}(\mathbf{x})e^{\ii(t+\ii\hbar \beta')\hat{H}^{\xi}/\hbar}\hat{j}_{\beta}(\mathbf{x'})\right]}\notag\\
&=-T\mathcal{L}\Bigg\{\int_{0}^{\beta}d\beta'\,\sum_{\lambda,\lambda'}\int \frac{d^2k_{\parallel}}{(2\pi)^2}\int \frac{d^2k_{\parallel}'}{(2\pi)^2}\notag\\
&\times \left\langle\Psi_{\lambda,\mathbf{k}_{\parallel}}\middle|\hat{\rho}_{0}e^{\ii(-t-\ii\hbar \beta')\hat{H}^{\xi}/\hbar}\hat{j}_{\alpha}(\mathbf{x})e^{\ii(t+\ii\hbar \beta')\hat{H}^{\xi}/\hbar}\middle|\Psi_{\lambda',\mathbf{k'}_{\parallel}}\middle\rangle\middle\langle\Psi_{\lambda',\mathbf{k'}_{\parallel}}\middle|\hat{j}_{\beta}(\mathbf{x'})\middle|\Psi_{\lambda,\mathbf{k}_{\parallel}}\right\rangle\Bigg\}\notag\\
&=-T\sum_{\lambda,\lambda'}\int \frac{d^2k_{\parallel}}{(2\pi)^2}\int \frac{d^2k_{\parallel}'}{(2\pi)^2}\,\,\int_{0}^{\infty}dt\,e^{-\left[\ii\left( \mathcal{E}^{\lambda,\xi}_{\mathbf{k}_{\parallel}}-\mathcal{E}^{\lambda',\xi}_{\mathbf{k'}_{\parallel}}\right)/\hbar+s\right]t}\int_{0}^{\beta}d\beta'e^{\left( \mathcal{E}^{\lambda,\xi}_{\mathbf{k}_{\parallel}}-\mathcal{E}^{\lambda',\xi}_{\mathbf{k'}_{\parallel}}\right)\beta'}\notag\\
&\quad\quad  \times \rho_0\left( \mathcal{E}^{\lambda,\xi}_{\mathbf{k}_{\parallel}}\right)\left\langle\Psi_{\lambda,\mathbf{k}_{\parallel}}\middle|\hat{j}_{\alpha}(\mathbf{x})\middle|\Psi_{\lambda',\mathbf{k'}_{\parallel}}\middle\rangle\middle\langle\Psi_{\lambda',\mathbf{k'}_{\parallel}}\middle|\hat{j}_{\beta}(\mathbf{x'})\middle|\Psi_{\lambda,\mathbf{k}_{\parallel}}\right\rangle\notag\\
&=-T\sum_{\lambda,\lambda'}\int \frac{d^2k_{\parallel}}{(2\pi)^2}\int \frac{d^2k_{\parallel}'}{(2\pi)^2}\,\,\frac{-1}{-\left[\ii\left( \mathcal{E}^{\lambda,\xi}_{\mathbf{k}_{\parallel}}-\mathcal{E}^{\lambda',\xi}_{\mathbf{k'}_{\parallel}}\right)/\hbar+s\right]}\frac{e^{\left( \mathcal{E}^{\lambda,\xi}_{\mathbf{k}_{\parallel}}-\mathcal{E}^{\lambda',\xi}_{\mathbf{k'}_{\parallel}}\right)\beta}-1}{\left( \mathcal{E}^{\lambda,\xi}_{\mathbf{k}_{\parallel}}-\mathcal{E}^{\lambda',\xi}_{\mathbf{k'}_{\parallel}}\right)}\notag\\
&\quad\quad  \times \frac{e^{-\beta \left(\mathcal{E}^{\lambda,\xi}_{\mathbf{k}_{\parallel}}-\mu\right)}}{\mathcal{Z}}\left\langle\Psi_{\lambda,\mathbf{k}_{\parallel}}\middle|\hat{j}_{\alpha}(\mathbf{x})\middle|\Psi_{\lambda',\mathbf{k'}_{\parallel}}\middle\rangle\middle\langle\Psi_{\lambda',\mathbf{k'}_{\parallel}}\middle|\hat{j}_{\beta}(\mathbf{x'})\middle|\Psi_{\lambda,\mathbf{k}_{\parallel}}\right\rangle.
\end{align}

Rearranging terms:
\begin{align}
    L_{\alpha\beta}^{(11)}(\mathbf{x},\mathbf{x'})&=-\hbar T\sum_{\lambda,\lambda'}\int \frac{d^2k_{\parallel}}{(2\pi)^2}\int \frac{d^2k_{\parallel}'}{(2\pi)^2}\left[\frac{-\ii\left( \mathcal{E}^{\lambda,\xi}_{\mathbf{k}_{\parallel}}-\mathcal{E}^{\lambda',\xi}_{\mathbf{k'}_{\parallel}}\right)+\hbar\,s}{\left( \mathcal{E}^{\lambda,\xi}_{\mathbf{k}_{\parallel}}-\mathcal{E}^{\lambda',\xi}_{\mathbf{k'}_{\parallel}}\right)^2+\hbar^2s^2}\right]\left[\frac{\rho_0\left( \mathcal{E}^{\lambda,\xi}_{\mathbf{k}_{\parallel}}\right)-\rho_0\left( \mathcal{E}^{\lambda',\xi}_{\mathbf{k'}_{\parallel}}\right)}{\left( \mathcal{E}^{\lambda,\xi}_{\mathbf{k}_{\parallel}}-\mathcal{E}^{\lambda',\xi}_{\mathbf{k'}_{\parallel}}\right)}\right]\notag\\
&\quad\quad  \times \left\langle\Psi_{\lambda,\mathbf{k}_{\parallel}}\middle|\hat{j}_{\alpha}(\mathbf{x})\middle|\Psi_{\lambda',\mathbf{k'}_{\parallel}}\middle\rangle\middle\langle\Psi_{\lambda',\mathbf{k'}_{\parallel}}\middle|\hat{j}_{\beta}(\mathbf{x'})\middle|\Psi_{\lambda,\mathbf{k}_{\parallel}}\right\rangle.
\end{align}

Note that the first term in the numerator of the first square bracket 
\bea
&&\frac{-\ii\left( \mathcal{E}^{\lambda,\xi}_{\mathbf{k}_{\parallel}}-\mathcal{E}^{\lambda',\xi}_{\mathbf{k'}_{\parallel}}\right)}{\left( \mathcal{E}^{\lambda,\xi}_{\mathbf{k}_{\parallel}}-\mathcal{E}^{\lambda',\xi}_{\mathbf{k'}_{\parallel}}\right)^2+\hbar^2s^2}\left[\frac{\rho_0\left( \mathcal{E}^{\lambda,\xi}_{\mathbf{k}_{\parallel}}\right)-\rho_0\left( \mathcal{E}^{\lambda',\xi}_{\mathbf{k'}_{\parallel}}\right)}{\left( \mathcal{E}^{\lambda,\xi}_{\mathbf{k}_{\parallel}}-\mathcal{E}^{\lambda',\xi}_{\mathbf{k'}_{\parallel}}\right)}\right]=-\ii\frac{\rho_0\left( \mathcal{E}^{\lambda,\xi}_{\mathbf{k}_{\parallel}}\right)-\rho_0\left( \mathcal{E}^{\lambda',\xi}_{\mathbf{k'}_{\parallel}}\right)}{\left( \mathcal{E}^{\lambda,\xi}_{\mathbf{k}_{\parallel}}-\mathcal{E}^{\lambda',\xi}_{\mathbf{k'}_{\parallel}}\right)^2+\hbar^2s^2}
\eea
\end{widetext}
is odd in the integration variables, and therefore, has zero contribution. As expected, no imaginary part survives. For the real part, we use
\begin{equation}
    \lim_{s\rightarrow 0^{+}}\frac{\hbar\,s}{\left( \mathcal{E}^{\lambda,\xi}_{\mathbf{k}_{\parallel}}-\mathcal{E}^{\lambda',\xi}_{\mathbf{k'}_{\parallel}}\right)^2+\hbar^2s^2}=\pi \delta \left(\mathcal{E}^{\lambda,\xi}_{\mathbf{k}_{\parallel}}-\mathcal{E}^{\lambda',\xi}_{\mathbf{k'}_{\parallel}}\right),
\end{equation}
so that for the term in the second square bracket
\bea
    &&\left[\frac{\rho_0\left( \mathcal{E}^{\lambda,\xi}_{\mathbf{k}_{\parallel}}\right)-\rho_0\left( \mathcal{E}^{\lambda',\xi}_{\mathbf{k'}_{\parallel}}\right)}{\left( \mathcal{E}^{\lambda,\xi}_{\mathbf{k}_{\parallel}}-\mathcal{E}^{\lambda',\xi}_{\mathbf{k'}_{\parallel}}\right)}\right]\delta \left(\mathcal{E}^{\lambda,\xi}_{\mathbf{k}_{\parallel}}-\mathcal{E}^{\lambda',\xi}_{\mathbf{k'}_{\parallel}}\right)\nn\\
    &=&\left.\frac{\partial f_0(E)}{\partial E}\right|_{E=\mathcal{E}^{\lambda,\xi}_{\mathbf{k}_{\parallel}}}\delta \left(\mathcal{E}^{\lambda,\xi}_{\mathbf{k}_{\parallel}}-\mathcal{E}^{\lambda',\xi}_{\mathbf{k'}_{\parallel}}\right),
\eea
where $f_0(E,T) = \left( 1 + \exp[\left(E - \mu  \right)/k_{B}T]  \right)^{-1}$ is the Fermi-Dirac distribution. Then 
  \begin{align}
    &L_{\alpha\beta}^{(11)}(\mathbf{x},\mathbf{x'})=-\pi \hbar v_\text{F}^2 T \nn\\
    &\times\sum_{\lambda,\lambda'}\int \frac{d^2k_{\parallel}}{(2\pi)^2}\int \frac{d^2k_{\parallel}'}{(2\pi)^2}\left.\frac{\partial f_0(E)}{\partial E}\right|_{E=\mathcal{E}^{\lambda,\xi}_{\mathbf{k}_{\parallel}}}\delta \left(\mathcal{E}^{\lambda,\xi}_{\mathbf{k}_{\parallel}}-\mathcal{E}^{\lambda',\xi}_{\mathbf{k'}_{\parallel}}\right)\nn\\
    &\times \Psi_{\lambda,\mathbf{k}_{\parallel}}^{\dagger}(\mathbf{x})\sigma_{\alpha} \Psi_{\lambda',\mathbf{k'}_{\parallel}}(\mathbf{x})\Psi_{\lambda',\mathbf{k'}_{\parallel}}^{\dagger}(\mathbf{x'})\sigma_{\beta}\Psi_{\lambda,\mathbf{k}_{\parallel}}(\mathbf{x'}),
\end{align} 
where we have used the representation of the current operator in Eq.~\eqref{eq:electric_current_op}. On the other hand, we know that
\begin{align}
&\Psi_{\lambda,\mathbf{k}_{\parallel}}^{\dagger}(\mathbf{x})\sigma_{\alpha} \Psi_{\lambda',\mathbf{k'}_{\parallel}}(\mathbf{x})\Psi_{\lambda',\mathbf{k'}_{\parallel}}^{\dagger}(\mathbf{x'})\sigma_{\beta}\Psi_{\lambda,\mathbf{k}_{\parallel}}(\mathbf{x'})\nn\\
&=\Tr\left[ \sigma_{\alpha}\Psi_{\lambda',\mathbf{k'}_{\parallel}}(\mathbf{x})\otimes\Psi_{\lambda',\mathbf{k'}_{\parallel}}^{\dagger}(\mathbf{x'})\sigma_{\beta} \Psi_{\lambda,\mathbf{k}_{\parallel}}(\mathbf{x'})\otimes\Psi_{\lambda,\mathbf{k}_{\parallel}}^{\dagger}(\mathbf{x})\right],
	\end{align}
and by defining the spectral function $\boldsymbol{\mathcal{A}}^{\xi}(\mathbf{x},\mathbf{x'};E)$:
\begin{eqnarray}
&&\boldsymbol{\mathcal{A}}^{\xi}(\mathbf{x},\mathbf{x'};E)\nn\\
&=& 2\pi\sum_{\lambda}\int\frac{d^2k_{\parallel}}{(2\pi)^2}
\Psi_{\lambda,\mathbf{k}_{\parallel}}(\mathbf{x})\otimes \Psi_{\lambda,\mathbf{k}_{\parallel}}^{\dagger}(\mathbf{x'} )\delta\left(E-\mathcal{E}^{\lambda,\xi}_{\mathbf{k}_{\parallel}} \right),\nn\\
\end{eqnarray}
together with the identity
\begin{align}
		&\int\frac{d^3k}{(2\pi)^3}\sum_{\lambda}g\left( \mathcal{E}^{\lambda,\xi}_{\mathbf{k}_{\parallel}}\right) \Psi_{\lambda,\mathbf{k}_{\parallel}}(\mathbf{x})\otimes \Psi_{\lambda,\mathbf{k}_{\parallel}}^{\dagger}(\mathbf{x'} )\notag\\
		&=\int_{-\infty}^{\infty}\frac{dE}{2\pi}\boldsymbol{\mathcal{A}}^{\xi}(\mathbf{x},\mathbf{x'};E)g\left( E\right), 
	\end{align}
we arrive at the result
{\small
\begin{align}
    &L_{\alpha\beta}^{(11)}(\mathbf{x},\mathbf{x'})\nn\\
    &=-\frac{\hbar v_\text{F}^2 T}{2\pi}\int_{-\infty}^{\infty}dE\,\,\frac{\partial f_0(E)}{\partial E}\,\Tr\left[\sigma_{\alpha}\boldsymbol{\mathcal{A}}^{\xi}(\mathbf{x},\mathbf{x'};E)\sigma_{\beta}\boldsymbol{\mathcal{A}}^{\xi}(\mathbf{x'},\mathbf{x};E)\right],\label{eq:L_11_rr'}
\end{align}
}
where an additional factor of 2 comes from the spin degeneracy. 

 The other Onsager's coefficients are obtained by the same fashion, so that:

 \begin{align}
    &L_{\alpha\beta}^{(12)}(\mathbf{x},\mathbf{x'})= L_{\alpha\beta}^{(21)}(\mathbf{x},\mathbf{x'})\nn\\
    &=-\frac{\hbar v_\text{F}^2 T}{2\pi}\int_{-\infty}^{\infty}dE\,\,\frac{\partial f_0(E)}{\partial E}\,(E-\mu)\nn\\
    &\times\Tr \left[\sigma_{\alpha}\boldsymbol{\mathcal{A}}^{\xi}(\mathbf{x},\mathbf{x'};E)\sigma_{\beta}\boldsymbol{\mathcal{A}}^{\xi}(\mathbf{x'},\mathbf{x};E)\right],\label{eq:L_12_rr'}
\end{align}
and 
\begin{align}
    L_{\alpha\beta}^{(22)}(\mathbf{x},\mathbf{x'})&= -\frac{\hbar v_\text{F}^2 T}{2\pi}\int_{-\infty}^{\infty}dE\,\,\frac{\partial f_0(E)}{\partial E}\,(E-\mu)^2\nn\\
    &\times\Tr \left[\sigma_{\alpha}\boldsymbol{\mathcal{A}}^{\xi}(\mathbf{x},\mathbf{x'};E)\sigma_{\beta}\boldsymbol{\mathcal{A}}^{\xi}(\mathbf{x'},\mathbf{x};E)\right].\label{eq:L_22_rr'}
\end{align}

Let us work on the spectral function. Its coordinate representation is given by
\begin{eqnarray}
	   &&\boldsymbol{\mathcal{A}}^{\xi}(\mathbf{x},\mathbf{x'};E)\nn\\
    &=&\int\frac{d^2k_{\parallel}}{(2\pi)^2}e^{i\mathbf{k}_{\parallel}\cdot(\mathbf{x}-\mathbf{x'})}\sum_{\lambda}\left( \sigma_0+\lambda\frac{\boldsymbol{\sigma\cdot}\mathbf{k}_{\parallel}}{|\mathbf{k}_\parallel|}\right) \mathcal{A}^{\lambda,\xi}(k_{\parallel};E),\nn\\
    \label{eq:spectral_coordinates}
\end{eqnarray}
so that by inserting Eqs.~\eqref{eq:L_11_rr'}-\eqref{eq:L_22_rr'} we have
\begin{eqnarray}
L_{\alpha\beta}^{(11)}(\mathbf{x},\mathbf{x}')
&=& \int \frac{d^2 q_{\parallel}}{(2\pi)^2}e^{i\mathbf{q}_{\parallel}\cdot\left( \mathbf{x} - \mathbf{x}' \right)}\nn\\
&\times&\left( \frac{\hbar v_\text{F}^2 T}{2\pi} \right)\int_{-\infty}^{\infty}dE \left( -\frac{\partial f_0 (E)}{\partial E} \right)\nonumber\\
&\times& \int \frac{d^2 k_{\parallel}}{(2\pi)^2}
\Tr \left[\sigma_{\alpha}\boldsymbol{\mathcal{A}}^{\xi}(\mathbf{k}_{\parallel} + \mathbf{q}_{\parallel};E)\sigma_{\beta}\boldsymbol{\mathcal{A}}^{\xi}(\mathbf{k}_{\parallel};E)\right]\nonumber\\
&=& \int \frac{d^2 q_{\parallel}}{(2\pi)^2}e^{i\mathbf{q}_{\parallel}\cdot\left( \mathbf{x} - \mathbf{x}' \right)}
L_{\alpha\beta}^{(11)}(\mathbf{q}_{\parallel};T).
\end{eqnarray}

Without loss of generality, we will define the Onsager's coefficients in the limit $\mathbf{q}_{\parallel}\rightarrow 0$. Then:
\begin{eqnarray}
L_{\alpha\beta}^{(11)}(T) &=& \lim_{\mathbf{q}_{\parallel}\rightarrow 0} L_{\alpha\beta}^{(11)}(\mathbf{q}_{\parallel};T)\nonumber\\
&=& \left( \frac{\hbar v_\text{F}^2 T}{2\pi} \right)\int_{-\infty}^{\infty}dE \left( -\frac{\partial f_0 (E)}{\partial E} \right)\nn\\
&\times&\int \frac{d^2 k_{\parallel}}{(2\pi)^2}\mathcal{A}^{\lambda,\xi}(k_{\parallel} ;E)\mathcal{A}^{\lambda,\xi}(k_{\parallel};E)\nonumber\\
&\times&\Tr \left[\sigma_{\alpha}\left( \sigma_0 + \lambda \frac{\boldsymbol{\sigma}\cdot\mathbf{k}_{\parallel}}{k_{\parallel}}
 \right)\sigma_{\beta}\left( \sigma_0 + \lambda \frac{\boldsymbol{\sigma}\cdot\mathbf{k}_{\parallel}}{k_{\parallel}}
 \right)\right].\nn\\
\end{eqnarray}

From the $SU(2)$ algebra, we can readily obtain the trace
{\small
\begin{eqnarray}
\Tr \left[\sigma_{\alpha}\left( \sigma_0 + \lambda \frac{\boldsymbol{\sigma}\cdot\mathbf{k}_{\parallel}}{k_{\parallel}}
 \right)\sigma_{\beta}\left( \sigma_0 + \lambda \frac{\boldsymbol{\sigma}\cdot\mathbf{k}_{\parallel}}{k_{\parallel}}
 \right)\right] = 4\frac{k_{\alpha}k_{\beta}}{k_{\parallel}^2},\nn\\
\end{eqnarray}
 }
 and hence we have
\begin{eqnarray}
L_{\alpha\beta}^{(11)}(T)&=& 4\left( \frac{\hbar v_\text{F}^2 T}{2\pi} \right) \int_{-\infty}^{\infty}dE \left( -\frac{\partial f_0 (E)}{\partial E} \right)\nn\\
&\times&\int \frac{d^2k_{\parallel}}{(2\pi)^2}\frac{k_{\alpha}k_{\beta}}{k_{\parallel}^2}\mathcal{A}^{\lambda,\xi}(k_{\parallel} ;E)\mathcal{A}^{\lambda,\xi}(k_{\parallel};E).\nn\\
\end{eqnarray}

Performing the angular integral over polar coordinates $d^2k_{\parallel} = d\phi dk_{\parallel}k_{\parallel}$ first, we get
\begin{eqnarray}
\int\frac{d^2 k_{\parallel}}{(2\pi)^2} \frac{k_{\alpha}k_{\beta}}{k_{\parallel}^2} =
\frac{\delta_{\alpha\beta}}{(2\pi)^2}\frac{\pi}{2}\int_{0}^{\infty}dk_{\parallel} \frac{\mathbf{k}_{\parallel}\cdot\mathbf{k}_{\parallel}}{k_{\parallel}}.
\end{eqnarray}

Finally, using the definition of the spectral function in terms of the retarded and advanced disorder-averaged Green's functions,
\begin{eqnarray}
\mathcal{A}^{\lambda,\xi}(k_{\parallel};E)
= \ii \left[ \langle G_{R}^{\lambda,\xi}(k_{\parallel};E)  \rangle - \langle G_{A}^{\lambda,\xi}(k_{\parallel};E)  \rangle\right],
\end{eqnarray}
we obtain (in the limit of low impurity concentrations $n_\text{imp} \ll 1$)
\begin{eqnarray}
L_{\alpha\beta}^{(11)}(T) &=& \delta_{\alpha\beta}4\pi\left( \frac{\hbar v_\text{F}^2 T}{\left(2\pi\right)^3} \right) \int_{-\infty}^{\infty}dE \left( -\frac{\partial f_0 (E)}{\partial E} \right)\nn\\
&\times&\int_{0}^{\infty} dk_{\parallel} \langle G_{R}^{\lambda,\xi}(k_{\parallel};E)  \rangle \langle G_{A}^{\lambda,\xi}(k_{\parallel};E)  \rangle \frac{\mathbf{k}_{\parallel}\cdot\mathbf{k}_{\parallel}}{k_{\parallel}}.\nonumber\\
\end{eqnarray}

\subsection{The Vertex Corrections and Relaxation Time}\label{sec:The Vertex Corrections and Relaxation Time}

To include the vertex corrections, as described in Ref.~\cite{nano12203711}, we can formally perform the substitution $\mathbf{k}_{\parallel}\rightarrow\mathbf{\Gamma}_\text{RA}(\mathbf{k}_{\parallel};E)$ for one of the momenta in the integral. Here, $\mathbf{\Gamma}_\text{RA}(\mathbf{k}_{\parallel};E)$ is the solution to the \textit{Bethe-Salpeter equation} depicted in the diagram Fig.~\ref{fig:bethe_salpeter}, given by
{\small
  \begin{align}
		&\boldsymbol{\Gamma}_\text{RA}(\mathbf{k}_{\parallel},E)=\mathbf{k}_{\parallel}\nn\\
  &+n_\text{imp} \int \frac{d^2k'_{\parallel}}{(2\pi)^2} \left\langle G_R^{\lambda,\xi}(\mathbf{k'}_{\parallel})\right\rangle\left\langle G_A^{\lambda,\xi}(\mathbf{k'}_{\parallel})\right\rangle\left| \hat{T}^{(\lambda,\xi)}_{\mathbf{k'}_{\parallel}\mathbf{k}_{\parallel}} \right|^2\boldsymbol{\Gamma}_\text{RA}(\mathbf{k'}_{\parallel},E). \label{eq:bethe_salpeter}
	    \end{align}
}
where $\hat{T}^{(\lambda,\xi)}_{\mathbf{k'}_{\parallel}\mathbf{k}_{\parallel}}$ is the $T$-matrix operator. 

Hence, we have
\begin{align} 
	   & L^{(11)}_{\alpha\beta}(T)=\delta_{\alpha\beta}\frac{\hbar v_{F}^2 T}{2\pi^2 } \int_{-\infty}^{\infty}dE\,\,\left(-\frac{\partial f_0(E)}{\partial E}\right)\nn\\
     &\times\int_{0}^{\infty} dk_{\parallel} \left\langle G_R^{\lambda,\xi}(\mathbf{k}_{\parallel})\right\rangle\left\langle G_A^{\lambda,\xi}(\mathbf{k}_{\parallel})\right\rangle\frac{\mathbf{k}_{\parallel}\cdot\boldsymbol{\Gamma}_\text{RA}(\mathbf{k}_{\parallel},E)}{k_{\parallel}} , \label{eq:L_11_T}	
	    \end{align}

     \begin{align}
	   &L^{(12)}_{\alpha\beta}(T)\notag\\&=\delta_{\alpha\beta}\frac{\hbar v_{F}^2 T}{2\pi^2 } \int_{-\infty}^{\infty}dE\,\,\left(-\frac{\partial f_0(E)}{\partial E}\right)(E - \mu)\nn\\ &\times\int_{0}^{\infty} dk_{\parallel} \left\langle G_R^{\lambda,\xi}(\mathbf{k}_{\parallel})\right\rangle\left\langle G_A^{\lambda,\xi}(\mathbf{k}_{\parallel})\right\rangle\frac{\mathbf{k}_{\parallel}\cdot\boldsymbol{\Gamma}_\text{RA}(\mathbf{k}_{\parallel},E)}{k_{\parallel}}
     , \label{eq:L_12_T}	
	    \end{align}
     and 
\begin{align}
	    &L^{(22)}_{\alpha\beta}(T)\notag\\&=\delta_{\alpha\beta}\frac{\hbar v_{F}^2 T}{2\pi^2 } \int_{-\infty}^{\infty}dE\,\,\left(-\frac{\partial f_0(E)}{\partial E}\right)(E - \mu)^2\nn\\ &\times\int_{0}^{\infty} dk_{\parallel} \left\langle G_R^{\lambda,\xi}(\mathbf{k}_{\parallel})\right\rangle\left\langle G_A^{\lambda,\xi}(\mathbf{k}_{\parallel})\right\rangle\frac{\mathbf{k}_{\parallel}\cdot\boldsymbol{\Gamma}_\text{RA}(\mathbf{k}_{\parallel},E)}{k_{\parallel}}. \label{eq:L_22_T}	
	    \end{align}

Now, the vertex function must be of the form $\boldsymbol{\Gamma}_\text{RA}(\mathbf{k}_{\parallel},E)=\gamma(\mathbf{k}_{\parallel},E)\mathbf{k}_{\parallel}$, with $\gamma(\mathbf{k}_{\parallel},E)$ a scalar function. Moreover, in the limit of low impurity concentration $n_\text{imp} \ll 1$ we have 
\begin{align}
	    \left\langle G_R^{\lambda,\xi}(\mathbf{k}_{\parallel})\right\rangle\left\langle G_A^{\lambda,\xi}(\mathbf{k}_{\parallel})\right\rangle& \rightarrow\frac{2\pi \tau(k_{\parallel})}{\hbar}\delta(E-\lambda\xi\hbar v_\text{F} k_{\parallel}). \label{eq:limit_GRGA}
	\end{align}

 Then, in such a limit, we obtain a secular integral equation for the scalar function $\gamma(\mathbf{k}_{\parallel},E)$
{\small
   	\begin{align}
&\gamma(\mathbf{k}_{\parallel},E)=1+n_\text{imp} \frac{2\pi}{\hbar}\int \frac{d^2k'_{\parallel}}{(2\pi)^2}\tau(k'_{\parallel})\nn\\
&\times\left| T^{(\lambda,\xi)}_{\mathbf{k'}_{\parallel}\mathbf{k}_{\parallel}} \right|^2\delta(E-\lambda\xi\hbar v_\text{F} k'_{\parallel})\,\,\gamma(\mathbf{k'}_{\parallel},E)\frac{\mathbf{k}_{\parallel}\cdot\mathbf{k'}_{\parallel}}{k^2_{\parallel}}. \label{eq:eqn_gamma}
	    \end{align}  
}

At low temperatures, an exact solution is possible since the derivative of the Fermi distribution takes a compact support at the Fermi energy. Therefore, we can evaluate $\gamma(k_{\parallel};E)$ and $\tau(k_{\parallel})$ at the Fermi momentum $k_\text{F}$, to obtain
\begin{equation}
	    \gamma(k_\text{F})=\frac{\tau_1(k_\text{F})}{\tau_1(k_\text{F})-\tau(k_\text{F})},
	    \label{eq_gamma_tau}
\end{equation}
 where we defined (for $\cos\phi'=\mathbf{k}_{\parallel}\cdot\mathbf{k'}_{\parallel}/k^2_{\parallel}$)
\bea
	    &&\frac{1}{\tau_1(k_\text{F})}\nn\\
     &=&\frac{2\pi n_\text{imp} }{\hbar}\int \frac{d^2k'_{\parallel}}{(2\pi)^2}\left| T^{(\lambda,\xi)}_{\mathbf{k'}_{\parallel}\mathbf{k}_{\parallel}} \right|^2\cos \phi'\,\,\delta(\hbar v_\text{F} k_\text{F}-\hbar v_\text{F} k'_{\parallel}).\nn\\ \label{eq:tau_1}
\eea

After the substitution of $\gamma(\mathbf{k}_{\parallel},E)$ from Eq.~\eqref{eq_gamma_tau}, and the simple relation
\begin{eqnarray}
-\frac{\partial f_0(E)}{\partial E} = \frac{1}{k_\text{B} T} f_0(E) \left[1 - f_0(E) \right]
\end{eqnarray}
we finally obtain for the bulk Onsager coefficients
	\begin{align}
	L^{(11)}_{\alpha\alpha}(T)&=\frac{ v_{F}^2}{\pi k_\text{B} }\tau_{\text{tr}}(k_\text{F})\int_{0}^{\infty}dk_{\parallel} k_{\parallel} \,  f_0\left(\mathcal{E}^{\lambda,\xi}_{\mathbf{k}_{\parallel}}\right) \left[1-f_0\left(\mathcal{E}^{\lambda,\xi}_{\mathbf{k}_{\parallel}}\right) \right],\label{eq:L_11_final}
	\end{align} 
  \begin{align}
	&L^{(12)}_{\alpha\alpha}(T)=\frac{ v_{F}^2}{\pi k_\text{B} }\tau_{\text{tr}}(k_\text{F})\nn\\
 &\times\int_{0}^{\infty}dk_{\parallel} k_{\parallel} \,  f_0\left(\mathcal{E}^{\lambda,\xi}_{\mathbf{k}_{\parallel}}\right) \left[1-f_0\left(\mathcal{E}^{\lambda,\xi}_{\mathbf{k}_{\parallel}}\right) \right](\lambda\xi\hbar v_\text{F} k_{\parallel}-\mu),\label{eq:L_12_final}
	\end{align} 
  and 
 \begin{align}
	&L^{(22)}_{\alpha\alpha}(T)=\frac{ v_{F}^2}{\pi k_\text{B} }\tau_{\text{tr}}(k_\text{F})\nn\\
 &\times\int_{0}^{\infty}dk \, k_{\parallel} \,  f_0\left(\mathcal{E}^{\lambda,\xi}_{\mathbf{k}_{\parallel}}\right) \left[1-f_0\left(\mathcal{E}^{\lambda,\xi}_{\mathbf{k}_{\parallel}}\right) \right](\lambda\xi\hbar v_\text{F} k_{\parallel}-\mu)^2.\label{eq:L_22_final}
	\end{align} 

By following Ref.~\cite{nano12203711} the total \textit{transport relaxation time} is defined by
	\begin{align}
	    &\frac{1}{\tau_{\text{tr}}(k_\text{F})}=\frac{1}{\tau(k_\text{F})}-\frac{1}{\tau_1(k_\text{F})}\\
	    &=\frac{2\pi n_\text{imp} }{\hbar}\int \frac{d^2k'}{(2\pi)^2}\delta(\hbar v_\text{F} k_\text{F}-\hbar v_\text{F} k')\left| T^{(\lambda,\xi)}_{\mathbf{k'}_{\parallel}\mathbf{k}_{\parallel}} \right|^2(1-\cos \phi'),\nonumber
	\end{align}
which can be expressed in terms of the scattering phase shifts $\delta_m(k)$
\begin{equation}
	    \frac{1}{\tau_{\text{tr}}(k_\text{F})}=\frac{2 n_\text{imp}  v_\text{F}}{k_\text{F}}\sum_{m=-\infty}^{\infty}\sin^2 \left[\delta_{m}(k_\text{F})-\delta_{m-1}(k_\text{F}) \right].
	    \label{eq_tau_k}
\end{equation}	

\subsection{The Electrical Conductivity}\label{sec:The Electrical Conductivity}

Note that $f_0(\varepsilon)\left[1-f_0(\varepsilon)\right]=f_0(\varepsilon)f_0(-\varepsilon)$ is an even function of its argument. Then we can write
\bea
    f_0\left(\mathcal{E}^{\lambda,\xi}_{\mathbf{k}_{\parallel}}\right) \left[1-f_0\left(\mathcal{E}^{\lambda,\xi}_{\mathbf{k}_{\parallel}}\right) \right]&=&\frac{e^{\lambda\xi(\hbar v_\text{F} k-\lambda\xi\mu)/k_\text{B} T}}{(e^{\lambda\xi(\hbar v_\text{F} k-\lambda\xi\mu)/k_\text{B} T}+1)^2}\nn\\
    &=&\frac{e^{(\hbar v_\text{F} k-\lambda\xi\mu)/k_\text{B} T}}{(e^{(\hbar v_\text{F} k-\lambda\xi\mu)/k_\text{B} T}+1)^2},\nn\\
\eea
provided by the fact that $\lambda\xi=\pm 1$. Let us introduce the variables
\begin{equation}
    x=\frac{\hbar v_\text{F}}{k_\text{B} T}k, \quad \tilde{\mu}=\frac{\lambda\xi\mu}{k_\text{B} T},\label{eq:new_varibles}
\end{equation}
so that
\begin{equation}
    f_0(x)(1-f_0(x))=\frac{e^{x-\tilde{\mu}}}{(e^{x-\tilde{\mu}}+1)^2}=\frac{\partial}{\partial \tilde{\mu}}\left(\frac{1}{e^{x-\tilde{\mu}}+1}\right).
\end{equation}

This change of variables implies that the desired integrals have the form
\begin{equation}
    I_n=\frac{d}{d\tilde{\mu}}\int_{0}^{\infty}\frac{x^n}{e^{x-\tilde{\mu}}+1}dx,~\text{for } n=2,3,4.
\end{equation}

By using the integral representation of the Polylogarithm function
\begin{equation}
    -\text{Li}_{s}(-z)=\frac{1}{\Gamma(s)}\int_{0}^{\infty}\frac{x^{s-1}}{e^x/z+1}dx,\label{eq:Li_def}
\end{equation}
and the derivative relation
\begin{equation}
    \frac{d}{d\tilde{\mu}}\text{Li}_{s}(-e^{\tilde{\mu}})=\text{Li}_{s-1}(-e^{\tilde{\mu}}),\label{eq:Li_deriv}
\end{equation}
one can show that the integral of interest is
\begin{equation}
    I_n=-n!\,\,\text{Li}_{n}(-e^{\tilde{\mu}}).\label{eq:I_n}
\end{equation}

Thus,
\begin{align}
    &L^{(11)(\lambda,\xi)}_{\alpha\alpha}(T)\nn\\
    &=\frac{ v_{F}^2}{\pi k_\text{B} }\tau_{\text{tr}}(k_\text{F})\int_{0}^{\infty}dk_{\parallel} \, k_{\parallel} \,  f_0\left(\mathcal{E}^{\lambda,\xi}_{\mathbf{k}_{\parallel}}\right) \left[1-f_0\left(\mathcal{E}^{\lambda,\xi}_{\mathbf{k}_{\parallel}}\right) \right]\notag\\
    &=\frac{ v_{F}^2}{\pi k_\text{B} }\tau_{\text{tr}}(k_\text{F})\left(\frac{k_\text{B} T}{\hbar v_\text{F}}\right)^2\frac{d}{d\tilde{\mu}}\int_{0}^{\infty}\frac{x}{e^{x-\tilde{\mu}}+1}dx\notag\\
    &=\frac{1}{\pi k_\text{B}}\left(\frac{k_\text{B} T}{\hbar }\right)^2\tau_{\text{tr}}(k_\text{F})\,\,I_1,\label{eq:L_11_I}
\end{align}
or by defining the chemical potential as $\mu = \hbar v_\text{F} k_\text{F}$
\begin{eqnarray}
     L^{(11)(\lambda,\xi)}_{\alpha\alpha}(T)&=& \frac{1}{\pi k_\text{B}}\left(\frac{k_\text{B} T}{\hbar }\right)^2\tau_{\text{tr}}(k_\text{F}) \ln\left(1 + e^{\frac{\lambda\xi\hbar v_\text{F} k_\text{F}}{k_\text{B} T}} \right).\nn\\
     \label{eq:L_11_final2}
\end{eqnarray}

The electrical conductivity is then obtained from Eq.\eqref{eq:conductivity_tensor}
\begin{align}
    \sigma_{\alpha\beta}&= \frac{e^2}{T}\sum_{\lambda=\pm1}\sum_{\xi=\pm1}L^{(11)(\lambda,\xi)}_{\alpha\alpha}(T)\notag\\
    &=\frac{2e^2}{T}\frac{1}{\pi k_\text{B}}\left(\frac{k_\text{B} T}{\hbar }\right)^2\tau_{\text{tr}}(k_\text{F})\nn\\
    &\times\left[ \ln\left(1 + e^{\frac{\hbar v_\text{F} k_\text{F}}{k_\text{B} T}} \right)+\ln\left(1 + e^{\frac{-\hbar v_\text{F} k_\text{F}}{k_\text{B} T}} \right)\right]\notag\\
    &=\frac{2e^2k_\text{B} T}{\pi \hbar^2}\tau_{\text{tr}}(k_\text{F})\left[ \ln\left(e^{\frac{\hbar v_\text{F} k_\text{F}}{k_\text{B} T}} \right)+2\ln\left(1 + e^{-\frac{\hbar v_\text{F} k_\text{F}}{k_\text{B} T}} \right)\right],
\end{align}
which after some elementary algebra reduces to Eq.~\eqref{eq:sigmaxx}.

\subsection{The Thermal Conductivity and Seebeck Coefficient}\label{sec:The Thermal Conductivity and Seebeck Coefficient}
By following the same procedure, the other coefficients are

 \begin{align}
    &L^{(12)(\lambda,\xi)}_{\alpha\alpha}(T)=\lambda\xi\frac{ v_{F}^2}{\pi k_\text{B} }\tau_{\text{tr}}(k_\text{F})\nn\\
    &\times\int_{0}^{\infty}dk_{\parallel} \, k_{\parallel} \,  f_0\left(\mathcal{E}^{\lambda,\xi}_{\mathbf{k}_{\parallel}}\right) \left[1-f_0\left(\mathcal{E}^{\lambda,\xi}_{\mathbf{k}_{\parallel}}\right) \right](\hbar v_\text{F} k_{\parallel}-\tilde{\mu}\cdot k_\text{B} T)\notag\\
    &=\lambda\xi\frac{ v_{F}^2}{\pi k_\text{B} }\tau_{\text{tr}}(k_\text{F})\left(\frac{k_\text{B} T}{\hbar v_\text{F}}\right)^2 (k_\text{B} T)\nn\\
    &\times\left[ \frac{d}{d\tilde{\mu}}\int_{0}^{\infty}\frac{x^2}{e^{x-\tilde{\mu}}+1}dx-\tilde{\mu}\frac{d}{d\tilde{\mu}}\int_{0}^{\infty}\frac{x}{e^{x-\tilde{\mu}}+1}dx\right]\notag\\
    &=\frac{\lambda\xi}{\pi }\frac{\hbar}{k_\text{B}}\left(\frac{k_\text{B} T}{\hbar }\right)^3\tau_{\text{tr}}(k_\text{F})\left[ I_2-\tilde{\mu}I_1\right]\label{eq:L_12_I},
\end{align}
and
\begin{align}
    &L^{(22)(\lambda,\xi)}_{\alpha\alpha}(T)=\frac{ v_{F}^2}{\pi k_\text{B} }\tau_{\text{tr}}(k_\text{F})\nn\\
    &\times\int_{0}^{\infty}dk_{\parallel} \, k_{\parallel}\,  f_0\left(\mathcal{E}^{\lambda,\xi}_{\mathbf{k}_{\parallel}}\right) \left[1-f_0\left(\mathcal{E}^{\lambda,\xi}_{\mathbf{k}_{\parallel}}\right) \right](\hbar v_\text{F} k_{\parallel}-\tilde{\mu}\cdot k_\text{B} T)^2\notag\\
    &=\frac{ v_{F}^2}{\pi k_\text{B} }\tau_{\text{tr}}(k_\text{F})\left(\frac{k_\text{B} T}{\hbar v_\text{F}}\right)^2(k_\text{B} T)^2\nn\\
    &\times\left[ \frac{d}{d\tilde{\mu}}\int_{0}^{\infty}\frac{x^3}{e^{x-\tilde{\mu}}+1}dx-2 \tilde{\mu}\frac{d}{d\tilde{\mu}}\int_{0}^{\infty}\frac{x^2}{e^{x-\tilde{\mu}}+1}dx\right.\nn\\
    &+\left. \tilde{\mu}^2\frac{d}{d\tilde{\mu}}\int_{0}^{\infty}\frac{x}{e^{x-\tilde{\mu}}+1}dx \right]\notag\\
    &=\frac{\hbar^2}{\pi k_\text{B}}\left(\frac{k_\text{B} T}{\hbar }\right)^4\tau_{\text{tr}}(k_\text{F})\left[I_3-2 \tilde{\mu}\,I_2+\tilde{\mu}^2\,I_1\right],\label{eq:L_22_I}
\end{align}
so that
\begin{align}
    & L^{(12)(\lambda,\xi)}_{\alpha\alpha}(T)\notag\\
    &=-\frac{2\lambda\xi}{\pi }\frac{\hbar}{k_\text{B}}\left(\frac{k_\text{B} T}{\hbar }\right)^3\tau_{\text{tr}}(k_\text{F})\nn\\
    &\times\left[ \,\text{Li}_2\left(-e^{\frac{\lambda\xi\hbar v_\text{F} k_\text{F}}{k_\text{B} T}}\right) -\frac{\lambda\xi\hbar v_\text{F} k_\text{F}}{k_\text{B} T} \,\text{Li}_1\left(-e^{\frac{\lambda\xi\hbar v_\text{F} k_\text{F}}{k_\text{B} T}}\right)\right], \label{eq:L_12_final2}
\end{align}
and
\begin{align}
  &L^{(22)(\lambda,\xi)}_{\alpha\alpha}(T)=-\frac{2\hbar^2}{\pi k_\text{B}}\left(\frac{k_\text{B} T}{\hbar }\right)^4\tau_{\text{tr}}(k_\text{F})\nn\\
  &\times\left[3 \,\text{Li}_3\left(-e^{\frac{\lambda\xi\hbar v_\text{F} k_\text{F}}{k_\text{B} T}}\right) -2\lambda\xi\frac{\hbar v_\text{F} k_\text{F}}{k_\text{B} T} \,\text{Li}_2\left(-e^{\frac{\lambda\xi\hbar v_\text{F} k_\text{F}}{k_\text{B} T}}\right)\right.\notag\\
    &\left.+\frac{1}{2}\left(\frac{\hbar v_\text{F}k_\text{F}}{ k_\text{B} T}\right)^2\,\text{Li}_1\left(-e^{\frac{\lambda\xi\hbar v_\text{F} k_\text{F}}{k_\text{B} T}}\right)\right]. \label{eq:L_22_final2}
\end{align}

Then, the thermal conductivity is obtained by replacing Eqs.~\eqref{eq:L_11_I}, \eqref{eq:L_12_I} and \eqref{eq:L_22_I} into Eq.~\eqref{eq:thermal_conductivity_tensor}
\begin{align}
    \kappa_{\alpha\alpha}^{(\lambda\xi)}(T)&=\frac{1}{T^2}\left[L^{(22)(\lambda,\xi)}_{\alpha\alpha}(T)-\frac{L^{(12)(\lambda,\xi)}_{\alpha\alpha}(T)L^{(21)(\lambda,\xi)}_{\alpha\alpha}(T)}{L^{(11)(\lambda,\xi)}_{\alpha\alpha}(T)}\right]\notag\\
    &=\frac{\hbar^2}{\pi k_\text{B} T^2}\left(\frac{k_\text{B} T}{\hbar }\right)^4\tau_{\text{tr}}(k_\text{F})\left(I_3-\frac{I_2^2}{I_1}\right),
\end{align}
in such a way that the final result is
\begin{align}
    &\kappa_{\alpha\alpha}(T)=-\frac{2\hbar^2}{\pi k_\text{B} T^2}\left(\frac{k_\text{B} T}{\hbar }\right)^4\tau_{\text{tr}}(k_\text{F})\nn\\
    &\times\sum_{\lambda,\xi=\pm1}\left[3\,\text{Li}_3\left(-e^{\frac{\lambda\xi\hbar v_\text{F} k_\text{F}}{k_\text{B} T}}\right)+2\frac{\left[\text{Li}_2\left(-e^{\frac{\lambda\xi\hbar v_\text{F} k_\text{F}}{k_\text{B} T}}\right)\right]^2}{\ln\left(1 + e^{\frac{\lambda\xi\hbar v_\text{F} k_\text{F}}{k_\text{B} T}} \right)}\right].\label{eq:thermal_conductivity_final}
\end{align}

The Seebeck coefficient is obtained replacing Eqs.~\eqref{eq:L_11_I} and \eqref{eq:L_12_I} into Eq.~\eqref{eq:Seebeck}
\begin{align}
  S(T)&=\frac{1}{eT}\sum_{\lambda,\xi}\frac{L^{(12)(\lambda,\xi)}_{\alpha\alpha}(T)}{L^{(11)(\lambda,\xi)}_{\alpha\alpha}(T)}\nn\\
  &=-\frac{k_\text{B}}{e}\sum_{\lambda=\pm1}\sum_{\xi=\pm1}\left( \frac{2\lambda\xi\text{Li}_2\left(-e^{\frac{\lambda\xi\hbar v_\text{F} k_\text{F}}{k_\text{B} T}}\right)}{\ln\left(1 + e^{\frac{\lambda\xi\hbar v_\text{F} k_\text{F}}{k_\text{B} T}} \right)}+\frac{\hbar v_\text{F} k_\text{F}}{k_\text{B} T}\right).
\end{align}

\bibliography{GrapheneTI.bib}
\end{document}